%% file: Main_astroph.tex
\documentclass[twocolumn, twocolappendix]{aastex63}

\usepackage{soul}
\usepackage{natbib}

\usepackage{ulem}
\usepackage{amsmath}
\usepackage{graphicx}

\accepted{October 31, 2022}
\submitjournal{ApJ}

\shorttitle{A selection of H$\rm\alpha$ emitters at $z=2.1$--$2.5$}
\shortauthors{Terao et al.}

\begin{document}
\title{A selection of H$\rm\alpha$ emitters at $z=2.1$--$2.5$ using the $K_s$-band photometry of ZFOURGE}

\correspondingauthor{Yasunori Terao}
\email{yterao.astronomy@gmail.com}

\author{Yasunori Terao}
\affiliation{Institute of Astronomy, The University of Tokyo, Osawa 2-21-1, Mitaka, Tokyo 181-0015, Japan}

\author[0000-0001-5185-9876]{Lee R. Spitler}
\affiliation{School of Mathematical and Physical Sciences, Macquarie University, Sydney, NSW 2109, Australia}
\affiliation{Research Centre in Astronomy, Astrophysics \& Astrophotonics, Macquarie University, Sydney, NSW 2109, Australia}
\affiliation{Australian Astronomical Optics, Faculty of Science and Engineering, Macquarie University, Macquarie Park, NSW 2113, Australia}

\author{Kentaro Motohara}
\affiliation{Advanced Technology Center, National Astronomical Observatory of Japan, Osawa 2-21-1, Mitaka, Tokyo 181-8588, Japan}
\affiliation{Department of Astronomy, Graduate School of Science, The University of Tokyo, 7-3-1 Hongo, Bunkyo-ku, Tokyo 113-0033, Japan}

\author{Nuo Chen}
\affiliation{Department of Astronomy, Graduate School of Science, The University of Tokyo, 7-3-1 Hongo, Bunkyo-ku, Tokyo 113-0033, Japan}

\begin{abstract}
Large and less-biased samples of star-forming galaxies are essential to investigate galaxy evolution. H$\rm\alpha$ emission line is one of the most reliable tracers of star-forming galaxies because its strength is directly related to recent star formation. 
However, it is observationally expensive to construct large samples of H$\rm\alpha$ emitters by spectroscopic or narrow-band imaging survey at high-redshifts.
In this work, we demonstrate a method to extract H$\rm\alpha$ fluxes of galaxies at $z=2.1$--$2.5$ from $K_s$ broad-band photometry of ZFOURGE catalog.
Combined with 25--39 other filters, we estimate the emission line fluxes by SED fitting with stellar population models that incorporate emission-line strengths.
2005 galaxies are selected as H$\rm\alpha$ emitters by our method and their fluxes show good agreement with previous measurements in the literature.
On the other hand, there are more H$\rm\alpha$ luminous galaxies than previously reported.
The discrepancy can be explained by extended H$\rm\alpha$ profiles of massive galaxies and a luminosity dependence of dust attenuation, which are not taken into account in the previous work.
We also find that there are a large number of low-mass galaxies with much higher specific star formation rate (sSFR) than expected from the extrapolated star formation main sequence.
Such low-mass galaxies exhibit larger ratios between H$\rm\alpha$ and UV fluxes compared to more massive high sSFR galaxies.
This result implies that a ``starburst'' mode may differ among galaxies: low-mass galaxies appear to assemble their stellar mass via short-duration bursts while more massive galaxies tend to experience longer-duration ($>10\ \mathrm{Myr}$) bursts.
\end{abstract}

\input{intro_astroph.tex}
\input{data_astroph.tex}
\input{HAEs_astroph.tex}
\input{summary_astroph.tex}


\bibliographystyle{aasjournal}
\bibliography{ref_thesis}

\end{document}

%% file: intro_astroph.tex
\section{Introduction}
\label{sec:intro}
Since galaxies increase their masses through star formation, it is important to investigate properties of star-forming galaxies for understanding galaxy evolution.
How actively galaxies form stars at a given epoch is typically measured with the cosmic star formation rate density (SFRD), which is the star formation rate (SFR) per unit volume. This rate peaked at cosmological redshifts of $z\sim 2$--$3$ \citep{madau_cosmic_2014}.

A correlation between galaxy stellar mass and SFR is also an important statistical property of galaxy populations.
This so-called star formation main sequence (SFMS) has been observationally confirmed at least up to $z \sim 4$ \citep[e.g., ][]{tomczak_sfr-m*_2016}.
Its normalization increases with redshift, which indicates that galaxies at higher redshift tend to have higher SFR per stellar mass (specific star formation rate; sSFR) than in the local universe.
Thus measurements of the SFMS tell us how star formation activities in galaxies change through cosmic time.

Scatter around the SFMS can constrain galaxy evolution models because it likely reflects the dominant mechanism of star formation in galaxies.
For example, \citet{tacchella_confinement_2016} have suggested that observed small scatter ($\sim 0.3\ \mathrm{dex}$) can be explained by a ``self-regulated'' process, 
where SF in galaxies is governed by a balance between supply and depletion of gas due to inflows and star formation, respectively, rather than bursty events such as major mergers.
Other studies have also shown that the variations in sSFR can be explained by the time evolution of gas accretion rate onto the galaxies \citep{dutton_origin_2010,dekel_toy_2013}.
Hence the galaxies on the SFMS are expected to evolve along the tight correlation until they start to quench at the high-mass end.

\citet{sparre_starbursts_2017} found that simulated galaxies in the FIRE cosmological hydrodynamical simulation \citep{hopkins_galaxies_2014} with $\log{(M/M_\odot)}<9$ show larger scatter around the SFMS than more massive simulated galaxies.
They have attributed this finding to the increasing impact of the supernova feedback in galaxies with shallower gravitational potentials, where outflows caused by supernovae suppress star formation and some time later significant inflows result in a star formation burst.
Since different SFR indicators are sensitive to star formation operating over different timescales, \citet{sparre_starbursts_2017} point out  that the H$\alpha$/UV luminosity ratio is a useful metric for capturing time since a significant change in the star formation rate of a galaxy. Their simulated low-mass galaxies show relatively large scatter in the H$\alpha$/UV luminosity ratios, indicating a range of star formation timescales occurring in this galaxy population.
Interestingly, recent observations have found that galaxies both in the local universe and at $z>4$ with larger H$\alpha$/UV \citep{emami_closer_2019, faisst_recent_2019}.
However, these works had limited numbers as it is observationally expensive to construct large sample of high-redshift low-mass galaxies with both the H$\alpha$ and UV luminosities.

When measuring the cosmic SFRD and SFMS, a large survey volume is required to reduce effects of cosmic variance and put meaningful constraints on theories of galaxy formation. Furthermore it is important to use a relatively unbiased selection for the sample and obtain robust SFRs.
While there are various SFR indicators available, H$\alpha$ emission line is the most direct and reliable probe. It is produced by ionizing photons from the most massive and short-lived stars and sensitive to short-period fluctuations of star-formation activity.
In addition, \citet{oteo_nature_2015} have found that star-forming galaxies traced by H$\alpha$ emission lines are sensitive to a diverse range of galaxies at $z\sim 2$.
Therefore we can obtain less-biased sample of star-forming galaxies with robust SFR measurements by using H$\alpha$ emission line as a tracer.

There have been various studies to probe star-forming galaxies at $z>1$ with H$\alpha$ emission line by both spectroscopy and narrow-band imaging. 
For example, \citet{reddy_mosdef_2018} measured H$\alpha$ fluxes of galaxies at $1.4<z<3.8$ and investigated their properties in combination with other optical emission lines using spectroscopic data of the MOSDEF survey. \citet{sobral_stellar_2014} 
used narrow-band imaging data of HiZELS \citep{geach_hizels:_2008, sobral_hizels:_2009} to derive H$\alpha$ luminosity functions at four redshift bins of $z=0.40, 0.87, 1.47$ and $2.23$.

One limitation of spectroscopic sample is that due to their sensitivity limits, there are very limited number of observations for low-mass ($\log{(M/M_{\odot})}\lesssim$9.0) galaxies making it challenging to make statements about this class of galaxy \citep[e.g.,][]{Sanders_2021}.
An issue with narrow-band imaging surveys is that they generally lack information about dust attenuation.
One way to overcome these limitations is to use multi-wavelength broad-band photometry, which allows the identification of low-mass galaxies over relatively large survey volumes.
Indeed, several works have identified H$\alpha$ emission line at $z>4$ from color excesses in broad-band Spitzer/IRAC photometry  \citep[e.g.,][]{smit_evidence_2014,smit_high-precision_2015,stark_keck_2013,schaerer_impact_2009,faisst_coherent_2016,faisst_recent_2019}.
At lower redshift, 
\citet{Saito_2020} and \citet{Onodera_2020} have extracted H$\alpha$ fluxes at $0.3<z<2.0$ and [OIII] fluxes at $3.0<z<3.7$, respectively, from broad-band photometry in combination with SED fitting to multi-wavelength data. See also \citet{forrest_zfourge:_2018} who extracted H$\alpha$ fluxes from composite SEDs consisting of stacked broad-band photometry SEDs. 

In this work, we estimate H$\alpha$ emission line fluxes of galaxies at $z \sim 2.1$--$2.5$ from $K_s$ broad-band photometry using the multi-wavelength catalog from the ZFOURGE survey \citep{straatman_fourstar_2016}.
As fluxes measured in broad-band filters contain both continuum and emission lines, it is important to separate each contribution.
We estimate continuum fluxes by SED fitting with the multi-band photometry so that other properties of galaxies, such as stellar mass and amount of dust extinction, can be simultaneously estimated if the effects of emission lines are properly taken into account.

This paper is organized as follows. Section \ref{sec:data} describes the data and methodology, where H$\alpha$ fluxes extracted from broad-band photometry are compared to those from narrow-band imaging and spectroscopy.
In Section \ref{sec:HAE}, we investigate properties of H$\alpha$ emitters at $z=2.1$--2.5. Especially we discuss their H$\alpha$ luminosity function focusing on the most luminous emitters{} and the low-mass end of SFMS where no large spectroscopic sample exists.
We summarize our results in Section \ref{sec:summary}

In the following we adopt the AB magnitude system and a cosmology with $H_0=70\ \mathrm{km\ s^{-1}\ Mpc^{-1}}, \Omega_m=0.3$, and $\Omega_\Lambda=0.7$.

%% file: data_astroph.tex
\section{Data and method}
\label{sec:data}

\subsection{ZFOURGE catalog}
\label{sec:catalog}
We use a photometric catalog of the FourStar galaxy evolution survey (ZFOURGE; \citealt{straatman_fourstar_2016}) for our analysis. Observations of the ZFOURGE were conducted over 45 nights with the FourStar near-infrared camera on the Magellan telescope \citep{persson_fourstar:_2013}, targeting three legacy survey fields: COSMOS \citep{scoville_cosmic_2007}, UDS \citep{lawrence_ukirt_2007}, and CDFS \citep{giacconi_chandra_2002}.
The survey area is $\sim 407$ $\mathrm{arcmin^2}$ in total.
Since a number of ground-based and space observations have been conducted in these fields, there is significant public data so that the catalog covers from 0.8--8 $\mathrm{\mu m}$ with 26--40 filters.
The resulting well-sampled galaxy SEDs makes it possible to accurately derive properties of galaxies by SED fitting.
The Spitzer/IRAC at 3.6 and 4.5 $\mathrm{\mu m}$ bands
are key for estimate stellar masses of the galaxies by putting good constraints on rest-frame near-infrared regimes of the SEDs at $z>1$.
In addition, the Spitzer/MIPS 24 $\mathrm{\mu m}$ band data is available for all of the three fields, which can be used to estimate SFRs from infrared bolometric luminosities.

A unique advantage of ZFOURGE is the very deep $K_\mathrm{s}$-band imaging whose 5$\sigma$ limiting magnitude is $\geq 25.5$ across the fields, resulting in 80\% completeness for galaxies with $\log{(M/M_\odot)} \sim 9.0$ at $z=2$ \citep{straatman_fourstar_2016}.

Another unique characteristic of the ZFOURGE survey is its medium-band filters dividing $J$ and $H$-bands into three ($J_\mathrm{1}, J_\mathrm{2}, J_\mathrm{3}$) and two ($H_\mathrm{s}$, $H_\mathrm{l}$), respectively.
These filters efficiently sample the Balmer break (3636 \AA) 
and 4000 \AA break of galaxies at $1.5\lesssim z \lesssim 3.5$ and improve photo-$z$ accuracy in this redshift range.
\citet{straatman_fourstar_2016} have derived photo-$z$s from SED fitting by EAZY \citep{brammer_eazy:_2008} and compared them with publicly available spectroscopic redshifts.
They have found median $\sigma_z \sim 0.013$--$0.022$ in the three fields, where $\sigma_z \equiv 1.48\times |z_{\mathrm{phot}} - z_{\mathrm{spec}}|/(1+z_{\mathrm{spec}})$.

\subsection{Flux excess in the \texorpdfstring{$K_\mathrm{s}$}{Ks}-band}
\label{sec:flux_excess}
\begin{figure}
  \includegraphics[width=0.45\textwidth]{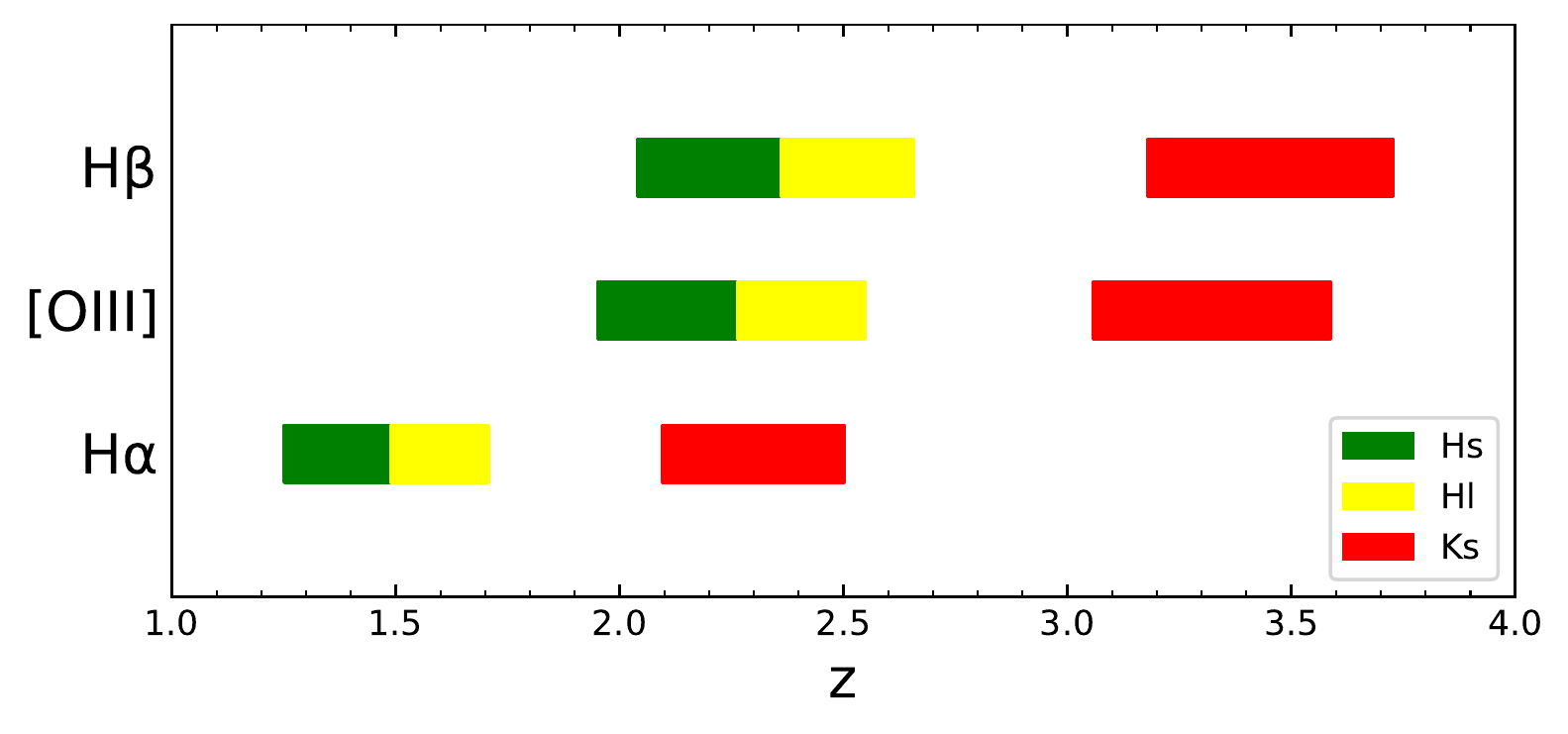}
  \caption{Combinations of the emission lines and the filters at each redshift.
  At $z>1$, H$\alpha$, H$\beta$, and [OIII] emission lines are observed in the near-infrared regime.
  Especially, at $2.1<z<2.5$, H$\alpha$ falls into the $K_\mathrm{s}$ broad-band filter of the ZFOURGE, while H$\beta$ and [OIII] are simultaneously observed in the $H$ medium-band filters.}
  \label{fig:line_and_filter}
\end{figure}

H$\alpha$ emission lines (rest-frame 6563 \AA) of galaxies at $2.1<z<2.5$ fall into the $K_\mathrm{s}$-band of the ZFOURGE, while [OIII] (rest-frame 4959 \& 5007 \AA) and H$\beta$ (rest-frame 4861 \AA) into the two medium-bands ($H_\mathrm{s}$ and $H_\mathrm{l}$), respectively as shown in \autoref{fig:line_and_filter}.

To derive the emission line fluxes, we define ``flux excess'' (in units of $\mathrm{erg\ s^{-1}\ cm^{-2}\ }$\AA$^{-1}$) as a difference between an observed flux and a continuum flux in a medium/broad-band filter as follows:
\begin{eqnarray}
\label{eq:flux_excess}
    f_{\textrm{excess}} = f_{\textrm{obs}} - f_{\textrm{cont}},
\end{eqnarray}
where $f_{\textrm{obs}}$ and $f_{\textrm{cont}}$ are the observed flux and the continuum flux, respectively.
The continuum flux is estimated by SED fitting taken into account of contributions of the emission lines.

\subsection{SED fitting with emission line templates}
\label{sec:SED}
When redshifted emission lines fall into certain filters, fluxes observed in them are boosted by the lines.
Hence the SED fitting may overestimate continuum fluxes of the galaxies unless the effect of the emission lines is properly taken into account.
This is especially important for high-$z$ galaxies because they have been known to show larger equivalent widths compared to those in the local universe \citep{stark_keck_2013,salmon_relation_2015,forrest_zfourge:_2018}.

Therefore, we run an SED fitting code, Fitting and Assessment of Synthetic Templates (FAST; \citealt{kriek_ultra-deep_2009}), with templates including emission lines to accurately estimate the continuum fluxes of the galaxies.
These templates are created by injecting emission line fluxes into a spectral templates of \citet{bruzual_stellar_2003} in a manner of \citet{salmon_relation_2015}\footnote{\url{https://github.com/BrettSalmon/Nebular}}.
They have calculated the emission-line strengths following a method of \citet{inoue_rest-frame_2011}, where strengths of the UV-optical emission lines from Ly$\alpha$ to those at $\sim 1\ \mathrm{\mu m}$ relative to H$\beta$ are calculated varying ionization parameters, metallicities, and hydrogen number densities using CLOUDY \citep{ferland_cloudy_1998}. 
H$\beta$ luminosity is estimated to be:
\begin{eqnarray}
    L_{\mathrm{H\beta}} = 4.78\times 10^{-13} \frac{1-f_{\mathrm{esc}}}{1+0.6f_{\mathrm{esc}}} N_{\mathrm{lyc}}
    \label{eq:line_strength}
\end{eqnarray}
where $f_{\mathrm{esc}}$ and $N_{\mathrm{lyc}}$ represent the escape fraction and the production rate of the Lyman continuum photons in a galaxy, respectively.
While $N_{\mathrm{lyc}}$ is calculated for each \cite{bruzual_stellar_2003} model template based on each age, there is no consensus on $f_{\mathrm{esc}}$ and its redshift dependency.
\citet{salmon_relation_2015} have provided calculations for two extreme cases, $f_{\mathrm{esc}}=0$ and $f_{\mathrm{esc}}=1$.
We adopt $f_{\mathrm{esc}}=0$, where all the Lyman continuum photons are absorbed and their energy is converted to the emission lines, to be consistent with a result obtained for Lyman break galaxies at $z\sim 3$ \citep{nestor_narrowband_2011}.

\begin{figure*}[p]
  \makebox[\textwidth][c]{\includegraphics[width=1.3\textwidth]{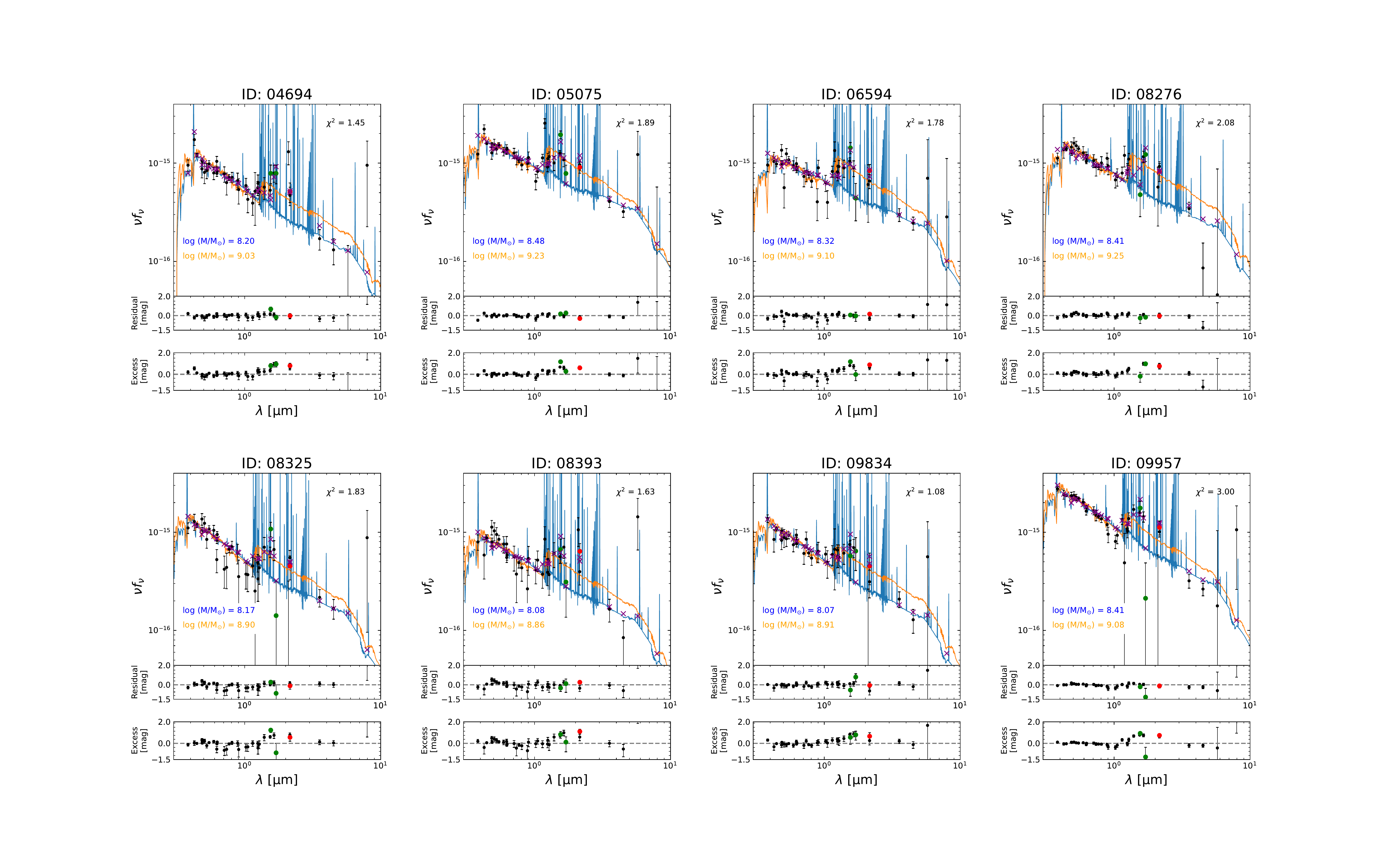}}
  \caption{Examples of the best-fit SEDs with (blue) and without (orange) the emission line templates, where IDs are from the ZFOURGE catalog.
  The Spitzer/IRAC photometry at 3.6 and 4.5 $\mathrm{\mu m}$ bands, which sample rest-frame $\sim 2\ \mathrm{\mu m}$, can not be fitted well unless the effects of the emission lines are taken into account.
  The estimated stellar masses can differ at most $\sim 0.8\ \mathrm{dex}$.
  The residuals shown just below the best-fit SEDs  are the differences between the observations and the best-fit templates including emission lines (purple crosses).
  On the other hand, the flux excesses shown in the bottom panels show the differences between the observations and the continuum fluxes estimated from the best-fit SEDs.
  Red circles represent the photometry of the $K_\mathrm{s}$-band, in which H$\alpha$ emission line falls.
  Green circles represent the H medium-band filters of ZFOURGE, in which H$\beta$ and/or O[III] emission lines fall.
  }
  \label{fig:SED}
\end{figure*}

\autoref{fig:SED} shows several examples of the best-fit SEDs obtained with and without the emission line templates (blue and orange, respectively).
It clearly shows that the continuum levels, which are mainly constrained by the Spitzer/IRAC photometry, can be significantly overestimated resulting in the overestimated stellar masses by at most $\sim 0.8\ \mathrm{dex}$.

\subsection{Robustness against model assumptions}
\label{sec:ad_excess}
\begin{figure*}[ht!]
  \makebox[\textwidth][c]{\includegraphics[width=0.9\textwidth]{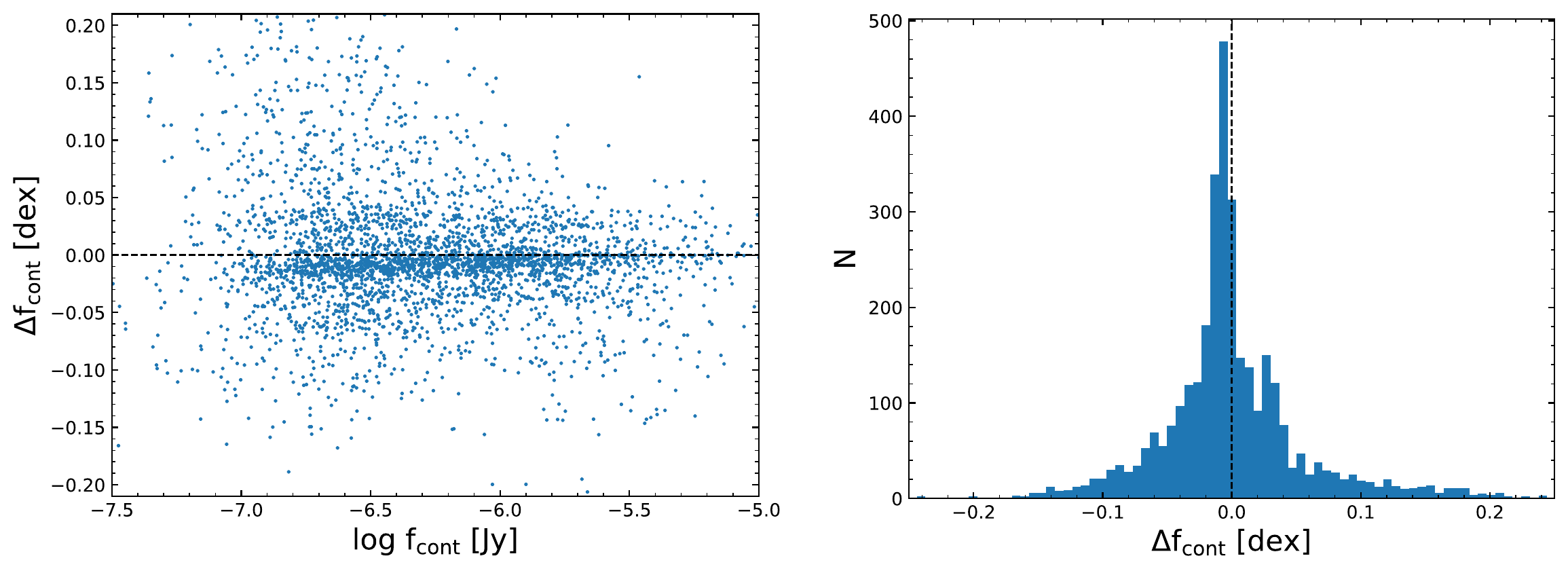}}
  \caption{(Left): Changes of the estimated continuum fluxes of our sample when assuming a doubled production rate of the Lyman continuum photons in the SED fitting.
  (Right): The distribution of the continuum offsets between the two models.
  }
  \label{fig:cont_flux}
\end{figure*}

The SED fitting described above allows us to derive not only the continuum levels but also the emission line strengths of galaxies.
However, the line strengths in the templates strongly depend on model assumptions, since the correct values of 
 $f_{\mathrm{esc}}$ and $N_{\mathrm{lyc}}$ are controversial and may vary among galaxies.
In contrast, the continuum fluxes estimated by the SED fitting are more robust against the model assumptions.
\autoref{fig:cont_flux} shows how the continuum fluxes of our sample vary when $N_{\mathrm{lyc}}$ is doubled in the SED fitting. Each panel shows flux difference due to $N_{\mathrm{lyc}}$ as a function of continuum flux (left) and a histogram (right).
We find a typical difference of $\lesssim 10\%$ between the continuum fluxes derived with the different assumptions.
Therefore we extract the continuum component from the SED fitting with the emission-line templates, and subtract them from observed fluxes to obtain H$\alpha$ fluxes.
In our sample, a typical difference between the H$\alpha$ fluxes from the flux excesses versus those derived from direct SED fitting is $\sim 40\%$.

\subsection{Selection of H\texorpdfstring{$\alpha$}{alpha} emitter at \texorpdfstring{$z=2.1$--$2.5$}{z=2.1--2.5}}
\label{sec:selection_HAE}
We select our initial candidates of H$\alpha$ emitters at $z=2.1$--$2.5$ by requiring the flux excesses in the $K_\mathrm{s}$-band to be $f_{\mathrm{excess}} \geq 2\sigma_{K_{\mathrm{s}}}$, where $\sigma_{K_\mathrm{s}}$ is photometric error in the band.
This selection results in 2005 H$\alpha$ emitters (699, 632, and 674 in COSMOS, UDS, and CDFS, respectively).

H$\alpha$ fluxes (in units of $\mathrm{erg\ s^{-1}\ cm^{-2}}$) of our sample are derived from the flux excesses as follows:
\begin{eqnarray}
    f_{\mathrm{H\alpha}} = r_{\mathrm{H\alpha}} \times f_{\mathrm{excess}} \times \Delta \lambda,
    \label{eq:f_ha}
\end{eqnarray}
where $r_{\mathrm{H\alpha}}$ is a ratio of H$\alpha$ to total strengths of all emission lines in the $K\mathrm{s}$-band, 
$f_{\mathrm{excess}}$ is the flux excess, and $\Delta \lambda$ is a bandwidth of the $K\mathrm{s}$ filter.
The other major emission lines expected to fall in the $K{\mathrm{s}}$-band are [NII] and [SII]. The constant $r_{\mathrm{H\alpha}}$ is calculated from the relative line-strengths used in Section \ref{sec:SED} adopting the photo-$z$ of individual galaxies to determine which lines fall into the $K\mathrm{s}$-band.
A median value of $r_{\mathrm{H\alpha}}$ is $\sim$0.79 over a metallicity range from $0.20Z_{\odot}$ to $Z_{\odot}$.
It increases up to $r_{\mathrm{H\alpha}} \sim 0.9$ for an extreme case of $Z=0.02Z_{\odot}$, however, the impact on the final $f_{\mathrm{H\alpha}}$ estimate is still insignificant.
These values are consistent with the line ratios of star-forming galaxies at $z\sim2$ measured by \citet{Kewley_2016} from spectroscopy.

Uncertainties of the H$\alpha$ fluxes are estimated using photometric errors in the $K{\mathrm{s}}$-band. The H$\alpha$ flux limit is calculated for each galaxy based on our flux excess threshold. 
As we will discuss in Section \ref{sec:HAE_EW}, our sample is mass-limited at $\log{(M/M_\odot)}>9.0$.

\subsection{H\texorpdfstring{$\alpha$}{alpha} flux comparisons with other methods}
\subsubsection{Narrow-band imaging}
The ZFOURGE catalog contains photometry of the {\it NB209} filter, which has been used in the New H$\alpha$ survey \citep{lee_dual-narrowband_2012}, for galaxies in COSMOS and CDFS.
The NB209 filter has been designed to have a central wavelength of 20,987 \AA\ and a FWHM of 208 \AA\ to detect H$\alpha$ from galaxies at $z=2.2$.

To check the consistency of our method with a conventional narrow-band emitter selection, we derive H$\alpha$ fluxes from $(K_\mathrm{s}-NB209)$ color excesses and compare them with those derived from the $K_\mathrm{s}$-band excesses.
Our selection procedure follows general techniques often used in narrow-band surveys
\citep[e.g.,][]{sobral_hizels:_2009,lee_dual-narrowband_2012}, which includes applying various S/N thresholds to identify high-quality flux measurements.
As a result of these quality filters, 237 galaxies are selected as emitter candidates.
We then narrow down the candidates to be at $2.1<z<2.5$ based on their photometric redshifts, which results in 73 H$\alpha$ emitters.

We find 63/73 (86\%) of the narrow-band selected galaxies are also selected by our method and the fraction does not depend on the amount of the color excess.
This suggests that our method can reproduce most of the narrow-band selected sample regardless of their equivalent widths.

\begin{figure}
  \includegraphics[width=0.5\textwidth]{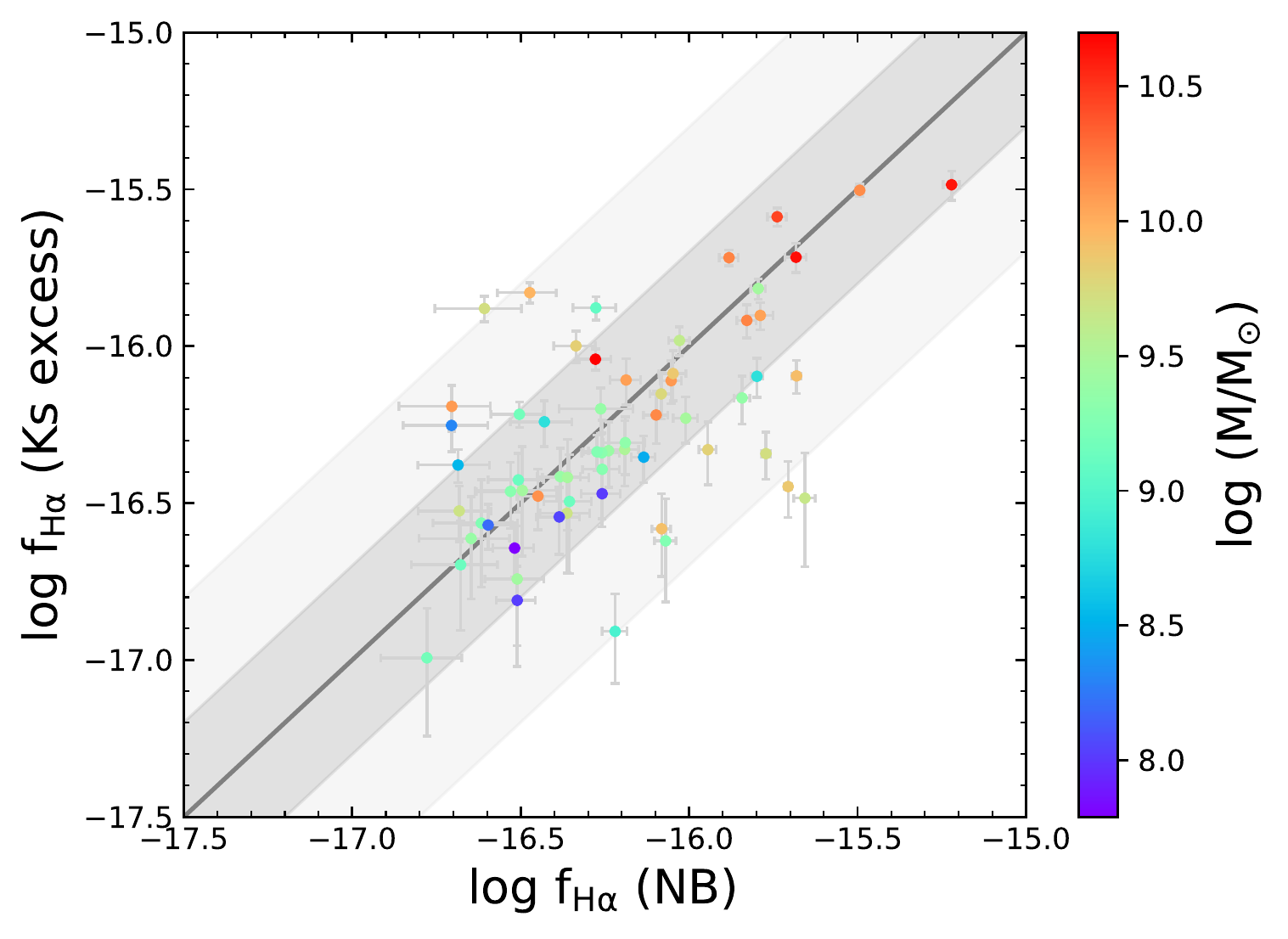}
  \caption{Comparison between the H$\alpha$ fluxes derived from the NB209 color excesses and the flux excesses in the $K_\mathrm{s}$-band.
  Both fluxes are in the units of $\mathrm{erg\ s^{-1}\ cm^{-2}}$.
  The grey line shows the one-to-one relation while the regions shaded in grey and light-grey indicate factor 2 and 5 intervals, respectively.
  The color coding of individual galaxies corresponds to their stellar masses.}
  \label{fig:fHa_vs_NB_wlogM}
\end{figure}

A comparison of the H$\alpha$ fluxes derived using the narrow-band color excesses and the $K_\mathrm{s}$-band excesses is shown in \autoref{fig:fHa_vs_NB_wlogM}.
We find that 46/63 (73\%) of the H$\alpha$ emitters have consistent values within a factor of 2 in both methods.
The typical scatter around the one-to-one relation is 0.16 dex.
Although some outliers from the one-to-one relation exist, there is no clear trend with stellar mass.

\subsubsection{Long-slit spectroscopy}
\label{sec:MOS}
\begin{figure}
  \includegraphics[width=0.5\textwidth]{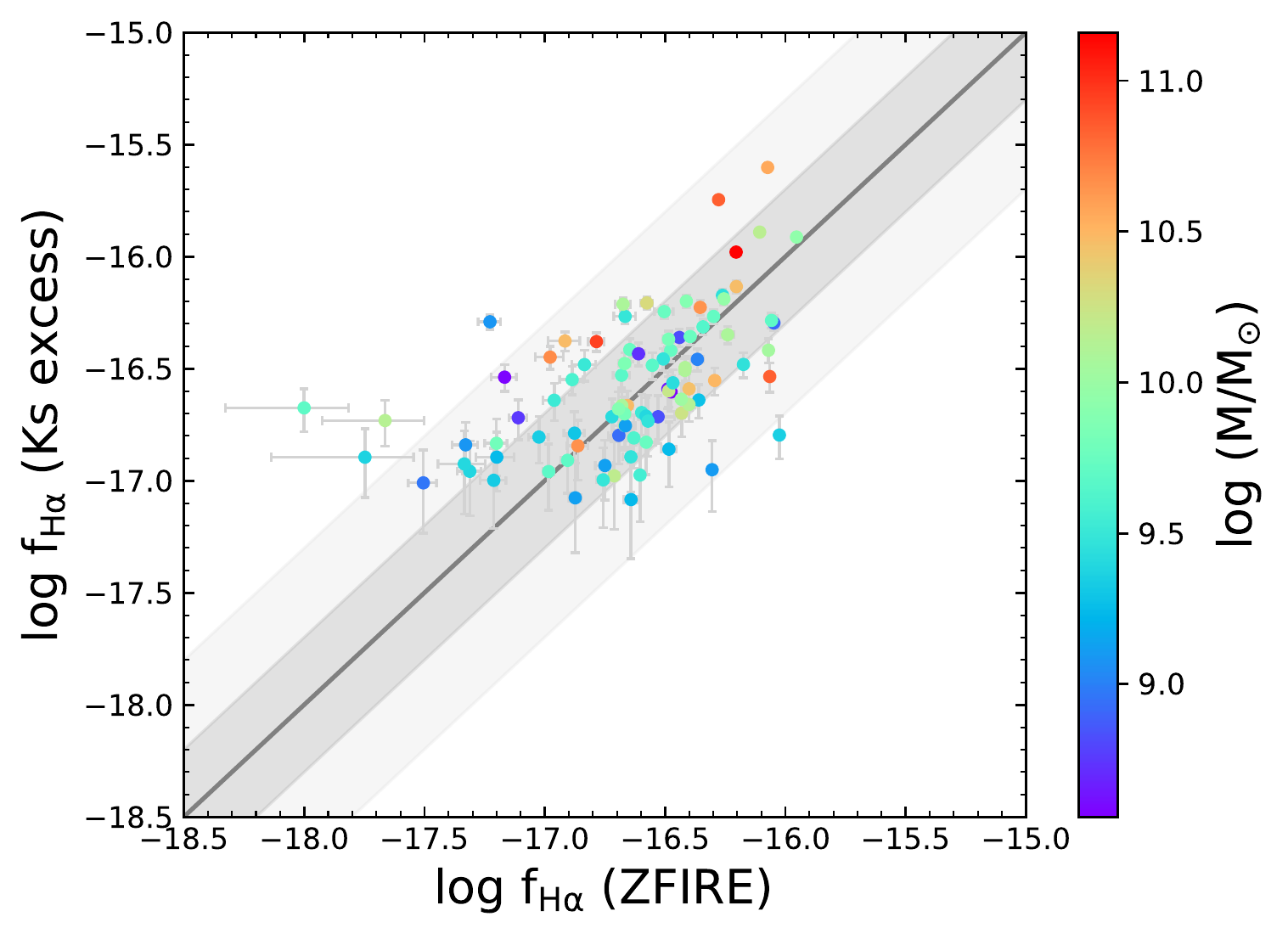}
  \caption{Same as \autoref{fig:fHa_vs_NB_wlogM}, but for the ZFIRE and our method, where the $K_\mathrm{s}$-band excesses are derived using the $1''.2$ aperture photometry.}
  \label{fig:fha_vs_ZFIRE_2}
\end{figure}

Some of the ZFOURGE galaxies have been observed by a spectroscopic survey, ZFIRE \citep{nanayakkara_zfire:_2016}. We find that 93 H$\alpha$ emitters in our sample have ZFIRE spectroscopic observations.
\autoref{fig:fha_vs_ZFIRE_2} shows a comparison of the H$\alpha$ fluxes between our method and ZFIRE, where a correction for different aperture sizes is applied.

We find that 62/93 (67\%) of the H$\alpha$ emitters have consistent fluxes in both methods within a factor of 2.
The typical scatter around the one-to-one relation is 0.21 dex.
Here the values of ZFIRE are  corrected for the slit loss assuming the aperture corrections applied to individual galaxies in the ZFOURGE, where the mean factor of the correction is $\sim 2.7$.

\subsubsection{Integral-field spectroscopy}
\begin{figure}
  \includegraphics[width=0.5\textwidth]{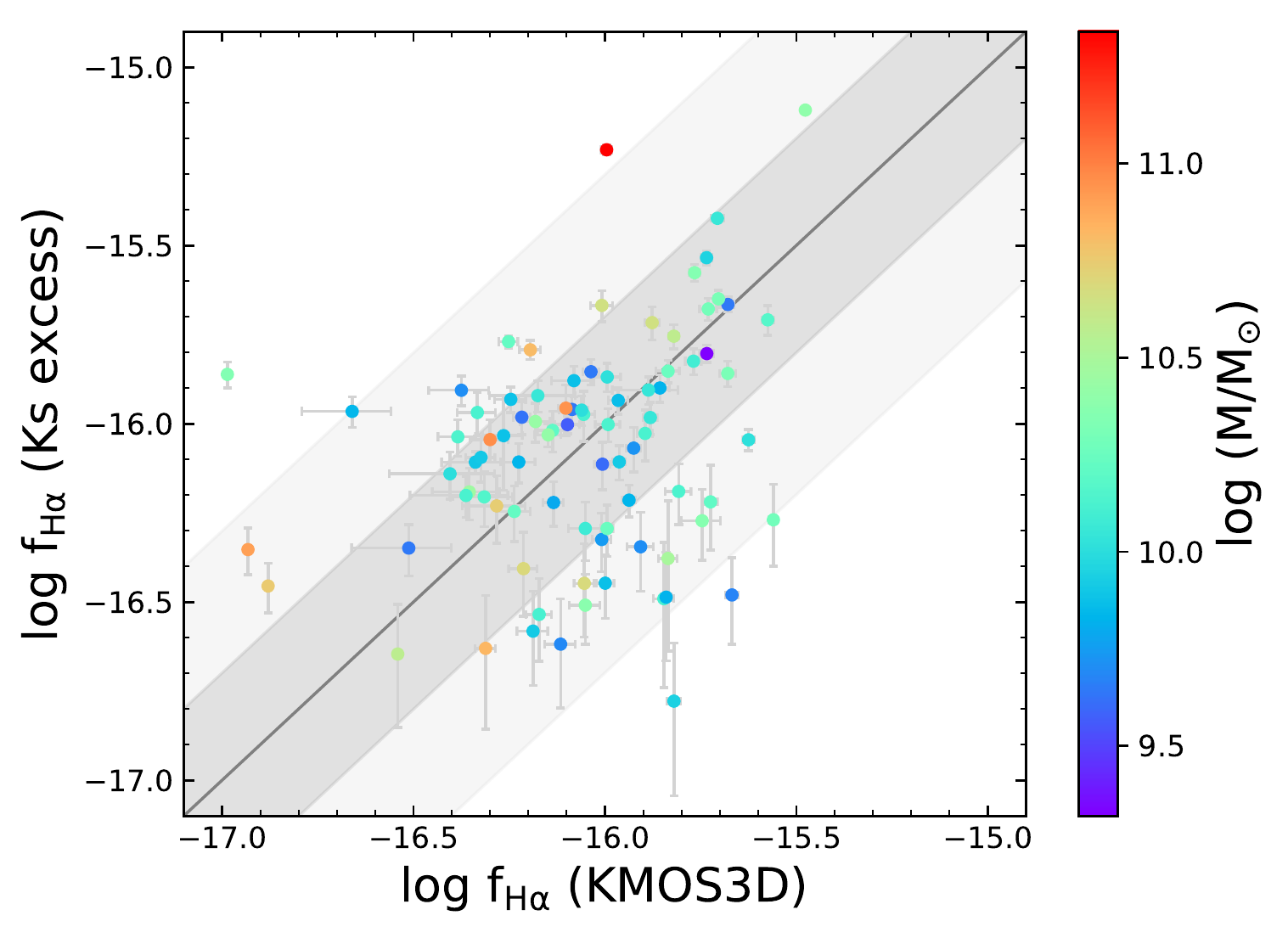}
  \caption{Same as \autoref{fig:fHa_vs_NB_wlogM}, but for the $\mathrm{KMOS^{3D}}$ and our method.}
  \label{fig:fha_vs_KMOS}
\end{figure}

We also compare our fluxes with the integral field spectroscopic (IFS) data taken by $\mathrm{KMOS^{3D}}$ survey \citep{wisnioski_kmos^3d_2019}.
In our sample, spectra of 86 H$\alpha$ emitters have been taken by the $\mathrm{KMOS^{3D}}$ (29, 21, and 36 in COSMOS, UDS, and CDFS, respectively).
\autoref{fig:fha_vs_KMOS} shows a comparison of the H$\alpha$ fluxes derived by our method and the IFS.
For 53/86 (62\%) of the H$\alpha$ emitters, the H$\alpha$ fluxes are consistent within a factor of 2.
The typical scatter around the one-to-one relation is 0.23 dex.

\subsubsection{Scatter between other methods}
As shown above, scatter around the one-to-one relations between our method and the narrow-band color excess, the long-slit spectroscopy, and the integral field spectroscopy are 0.16 dex, 0.21 dex, and 0.23 dex, respectively.
To check whether the scatter is caused by uncertainties in our method, we compare the H$\alpha$ fluxes between the three methods already reported in the literature.

The scatter between the ZFIRE and $\mathrm{KMOS^{3D}}$ is 0.16 dex, while the scatter between the ZFIRE and the narrow-band (NB209) color excess is 0.22 dex.
There are only four galaxies observed in both the NB209 and the $\mathrm{KMOS^{3D}}$.
We find a mean value of their flux differences is 0.23 dex, however it is difficult to extract statistical conclusion due to the small size of the sample.

Although the numbers of galaxies are small in the above comparisons, the scatters are similar to those found between our method and the others.
Therefore we conclude that there is no significant difference in the uncertainties of our H$\alpha$ flux derivation compared to other methods.

In addition, we have also applied our method to $H$ medium bands where [OIII] (4959, 5008 $\mathrm{\AA}$) and H$\beta$ emission lines of our H$\alpha$ emitters fall into.
The fluxes of the [OIII] doublet have been derived from the $H$-band excesses assuming a H$\alpha$/H$\beta$ ratio of 2.86. When compared to spectroscopic measurements of 3D-HST \citep[]{Brammer_2012,Momcheva_2016} they show good agreement with a typical scatter of $< 0.17\ \mathrm{dex}$ (N. Chen et al., in preparation).

\subsection{Corrections for dust extinction}
\label{sec:dust_correction}
When converting the H$\alpha$ fluxes to luminosities, we apply corrections for dust extinction.
According to \citet{nordon_far-infrared_2013}, the amount of dust extinction at rest-frame 1600 \AA\ is derived from infrared excess (IRX) as follows:
    \begin{eqnarray}
    \label{eq:IRX}
    A_{1600} = 2.5\log\left\{{\frac{\mathrm{SFR(IR)}}{\mathrm{SFR(UV)}}+1}\right\},
    \end{eqnarray}
where $\mathrm{SFR(IR)}$ and $\mathrm{SFR(UV)}$ are SFRs measured from IR luminosity and UV continuum luminosity, respectively.
The IR luminosity is estimated by integrating the IR spectral template of \citet{wuyts_what_2007} over 8--1000 $\mathrm{\mu m}$ scaled to match an observed 24 $\mathrm{\mu m}$ flux, while the UV luminosity is estimated 
from flux density at 1600 \AA\ obtained by SED fitting by EAZY \citep{brammer_eazy:_2008}, assuming the flatness of spectral slope of UV continuum between 1216--3000 \AA\ \citep{bell_toward_2005}.

We then  convert $A_{1600}$ to $A_{\mathrm{H\alpha}}$ assuming the extinction curve of \citet{calzetti_dust_2000}.
In this conversion, we assume that an $f$-factor, a ratio between dust extinctions of stellar continuum and nebular continuum to be unity.
Although there are uncertainties, it has been known that the $f$-factor increases with redshift and becomes nearly unity at $z\sim2$ \citep[e.g.,][]{erb_h_2006,koyama_predicting_2015,kashino_fmos-cosmos_2017}.
Moreover, \citet{faisst_recent_2019} have found that even in the local universe, where a typical $f$-factor is $\sim 0.44$, galaxies with high H$\alpha$ equivalent widths ($\gtrsim 100$ \AA) have $f$-factors close to unity. 
As the redshifts of our H$\alpha$ emitters are larger than 2 and their typical equivalent width is larger than 100 \AA\ (as we will see later), the assumption of $f$-factor to be unity for our sample is reasonable.

\begin{figure}
  \includegraphics[width=0.5\textwidth]{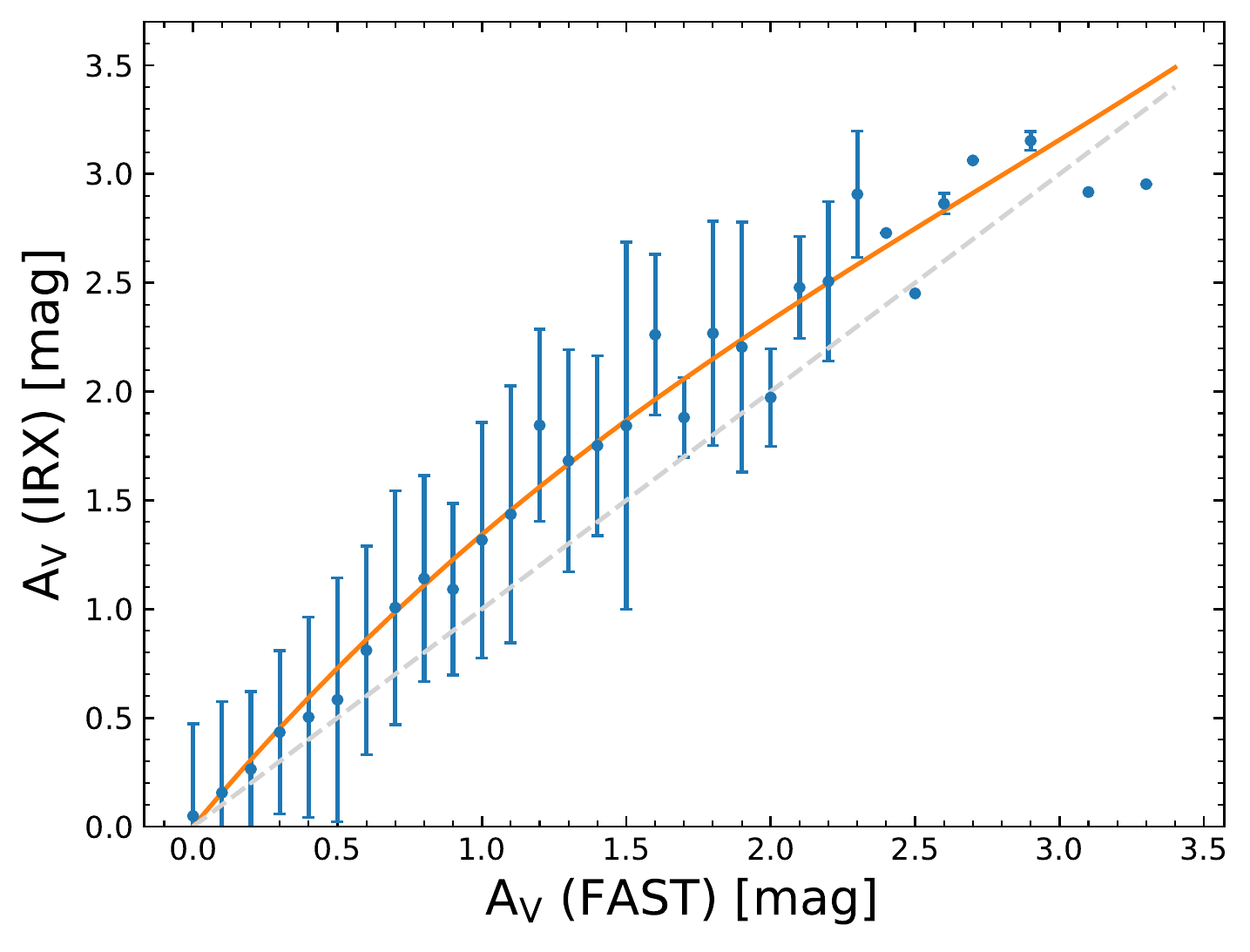}
  \caption{Relation between $A_{V}$ derived from the SED fitting by FAST and the IRX. Data are median values for each bin from all galaxies without a restriction on S/N. Orange curve shows the best-fit relation. Grey dashed line represent the one-to-one relation as a reference.
  }
  \label{fig:Av_relation}
\end{figure}

For galaxies without detections in the Spitzer/MIPS 24 $\mathrm{\mu m}$ band (S/N $<$ 3), we use relation between visual extinction derived from SED fitting by FAST ($A_{V,\rm{FAST}}$) and IRX extinction obtained from stacked MIPS 24$\mu$m fluxes ($A_{V,\mathrm{IRX}}$) as outlined in the following.
First, we divide our sample into $A_{V,\rm{FAST}}$ bins regardless of the S/N of the MIPS detections.
Median of 24 $\mathrm{\mu m}$ and 1600 \AA\ fluxes are calculated in each bin to obtain $A_{1600}$ via \autoref{eq:IRX}, which is then converted to $A_{V\rm, IRX}$ assuming the extinction curve of \citet{calzetti_dust_2000}.
\autoref{fig:Av_relation} shows a correlation between$A_{V,\rm{IRX}}$ and $A_{V,\rm{FAST}}$, where the best-fit relation is given by

\begin{equation}
\begin{split}
    A_{V,\mathrm{IRX}}=0.034 A_{V,\mathrm{FAST}}^3 - 0.230 A_{V,\mathrm{FAST}}^2 + 1.59 A_{V,\mathrm{FAST}}.
\end{split}
\end{equation}
We use this relation to estimate
visual extinctions for galaxies without IR detection, which is then converted to $A_{\mathrm{H\alpha}}$ using the curve of \citet{calzetti_dust_2000} again.

We find that there is an offset between $A_{V,\mathrm{IRX}}$ and $A_{V,\mathrm{FAST}}$.
For galaxies with $\mathrm{SFR(H\alpha)}>10$, a typical value of $A_{\mathrm{H\alpha}}$ based on the IRX is $\sim 0.21\ \mathrm{mag}$ larger than that from the SED fitting.
This discrepancy is consistent with the underlying physical tracers of dust, which differs between these methods. 
In the SED fitting, attenuation is estimated from a slope of the UV continua via an empirical relation \citep[IRX-$\beta$ relation; e.g.,][]{meurer_dust_1999,overzier_dust_2011}. If a galaxy is particularlly dusty, UV emission from star forming regions can be absorbed by dust and hence UV observations may not reflect the true star formation rate
\citep[e.g.,][]{qin_understanding_2019}.
In contrast, the IRX method indirectly captures the absorbed and re-radiated UV emission.
Therefore the IRX can better trace the total amounts of the attenuation by dust than the SED fitting which may underestimate the attenuation in the dusty galaxies. This is what we find in the observational data presented in \autoref{fig:Av_relation}.

We derive intrinsic H$\alpha$ luminosities corrected for dust extinction and convert them to H$\alpha$ SFRs using the calibration of \citet{kennicutt_star_1998} with a correction to the \citet{chabrier_galactic_2003} IMF:
\begin{eqnarray}
    L_{\mathrm{H\alpha}} &=& f_{\mathrm{H\alpha}}\times 4\pi d_{\mathrm{L}}^2 \times 10^{0.4A_{\mathrm{H\alpha}}}, \\
    \mathrm{SFR(H\alpha)} &=& 7.9 \times 10^{-42} L_{\mathrm{H\alpha}} \times 10^{-0.24},
\end{eqnarray}
where $d_{\mathrm{L}}$ is the luminosity distance corresponding to the redshift of a galaxy.

%% file: HAEs_astroph.tex
\section{Properties of H\texorpdfstring{$\alpha$}{alpha} emitters at \texorpdfstring{$z=2.1$--$2.5$}{z=2.1--2.5}}
\label{sec:HAE}

\subsection{Locations on the UVJ diagram}
\label{sec:UVJ_HAE}
\begin{figure}
  \centering
  \includegraphics[width=0.5\textwidth]{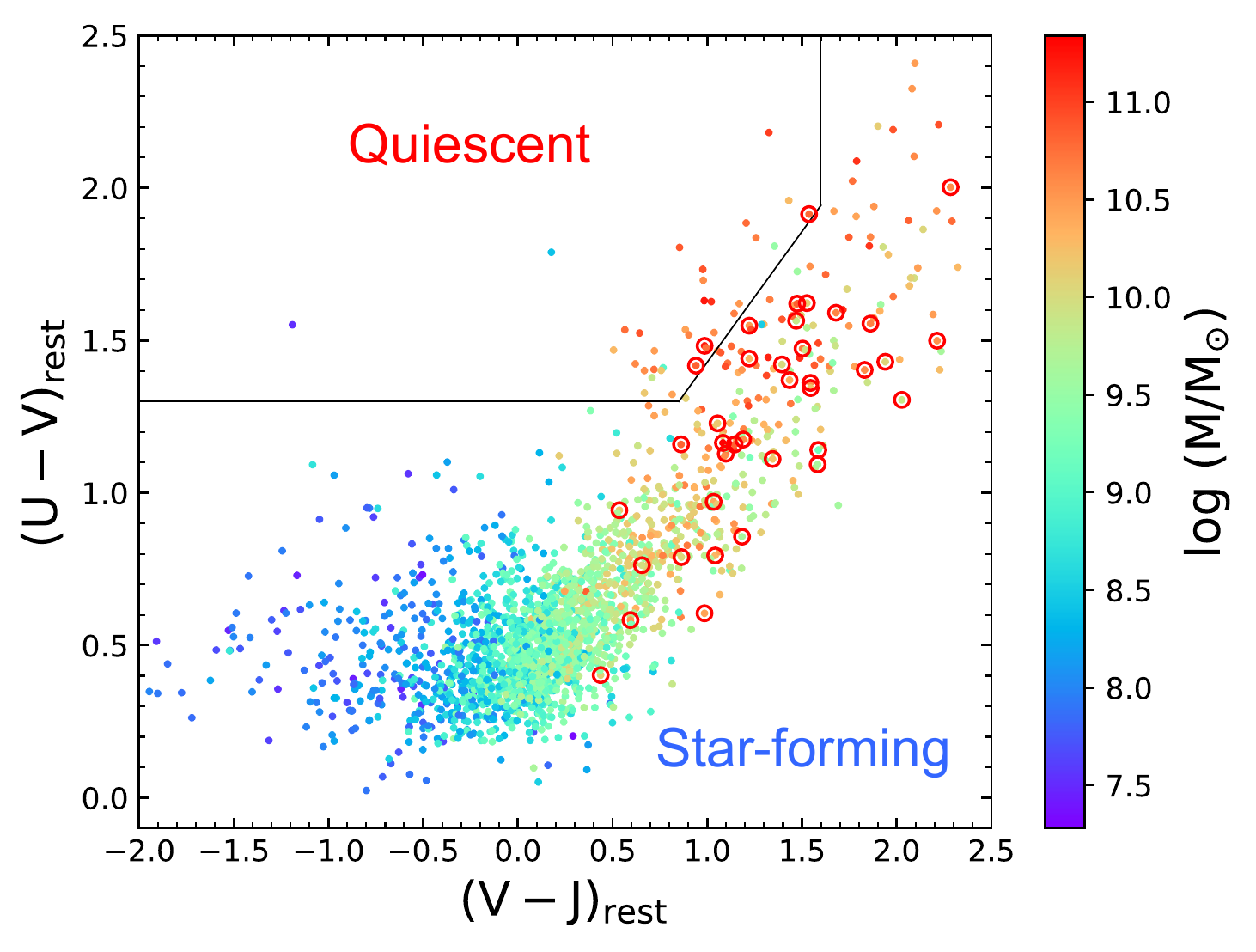}
  \caption{Locations of the H$\alpha$ emitters on the restframe $UVJ$ diagram \citep[e.g.][]{wuyts_what_2007}. Symbol color reflects galaxy stellar mass. 
  Red circles indicate objects identified as AGNs by either IR, radio, or X-ray in the catalog of \citet{cowley_zfourge_2016}.
Black line represents the criterion of \citet{spitler_exploring_2014} separating the star-forming and quiescent galaxies.
Identified AGNs by either IR, radio, or X-ray are indicated by the red open circles.}
  \label{fig:UVJ_HAE_logM}
\end{figure}

\autoref{fig:UVJ_HAE_logM} shows locations of the 2005 galaxies selected as the H$\alpha$ emitters on the restframe $UVJ$ diagram \citep[e.g.][]{wuyts_what_2007}. The restframe UVJ data is provided in the ZFOURGE catalogs.
It is clear that most of the H$\alpha$ emitters are classified as star-forming galaxies when we adopt the criterion of \citet{spitler_exploring_2014} for the classification.

Although the fraction is very small, there are 37 H$\alpha$ emitters in the quiescent region.
We note that a few low-mass galaxies with $M_*<10^{9} M_{\odot}$ (shown as bluer color in \autoref{fig:UVJ_HAE_logM}) classified as quiescent have poor SED fitting so that 
their stellar masses and H$\alpha$ excesses are likely not reliable.
However, most others have well-fitted SEDs and indeed a more conservative excess selection thresholds only reduces the H$\alpha$ emitters in the quiescent region from 37 to 22 with 3$\sigma_{K_{\mathrm{s}}}$ excess threshold.
One possibility is that our emitter selection is sensitive enough to detect weak H$\alpha$ emission lines from galaxies in quenching phases.
Indeed, the quiescent H$\alpha$ emitters show lower SFRs than star-forming galaxies with a similar-mass. As shown later below, they are also located below the star formation main sequence at this redshift range.

Another possibility is contributions from unidentified AGNs. As shown in \autoref{fig:UVJ_HAE_logM} there are a few known AGN in the quiescent region, which might mean some of these have emission-line contributions from the AGN.

Finally, it is possible that these galaxies are actually dusty star-forming galaxies who have scattered into the quiescent region due to measurement uncertainty in the $UVJ$ data. In any case, the existence of the quiescent H$\alpha$ emitters does not affect the results below because they only represent 1.8\% of the galaxies in H$\alpha$ luminosity range of $\log{L_{\rm{H\alpha}}}<43.2$.

\subsection{Deriving equivalent width}
\label{sec:HAE_EW}
A total rest-frame equivalent width of all emission lines in the $K_\mathrm{s}$-band is calculated as follows:
\begin{eqnarray}
\label{eq:limEW_all}
   EW(\mathrm{all})=\frac{f_{\mathrm{excess}}}{f_{\mathrm{cont}}} \times W / (1+z),
\end{eqnarray}
where $W$ is the bandwidth of the $K_\mathrm{s}$-band filter.
The H$\alpha$ equivalent width is obtained by:
\begin{eqnarray}
\label{eq:limEW_Ha}
   EW(\mathrm{H\alpha}) = r_{\mathrm{H\alpha}} \times EW(\mathrm{all}).
\end{eqnarray}
using the $r_{\mathrm{H\alpha}}$ ratio of $H\alpha$ to total strengths of all emission lines as described in \autoref{sec:selection_HAE}.

\begin{figure}
  \centering
  \includegraphics[width=0.5\textwidth]{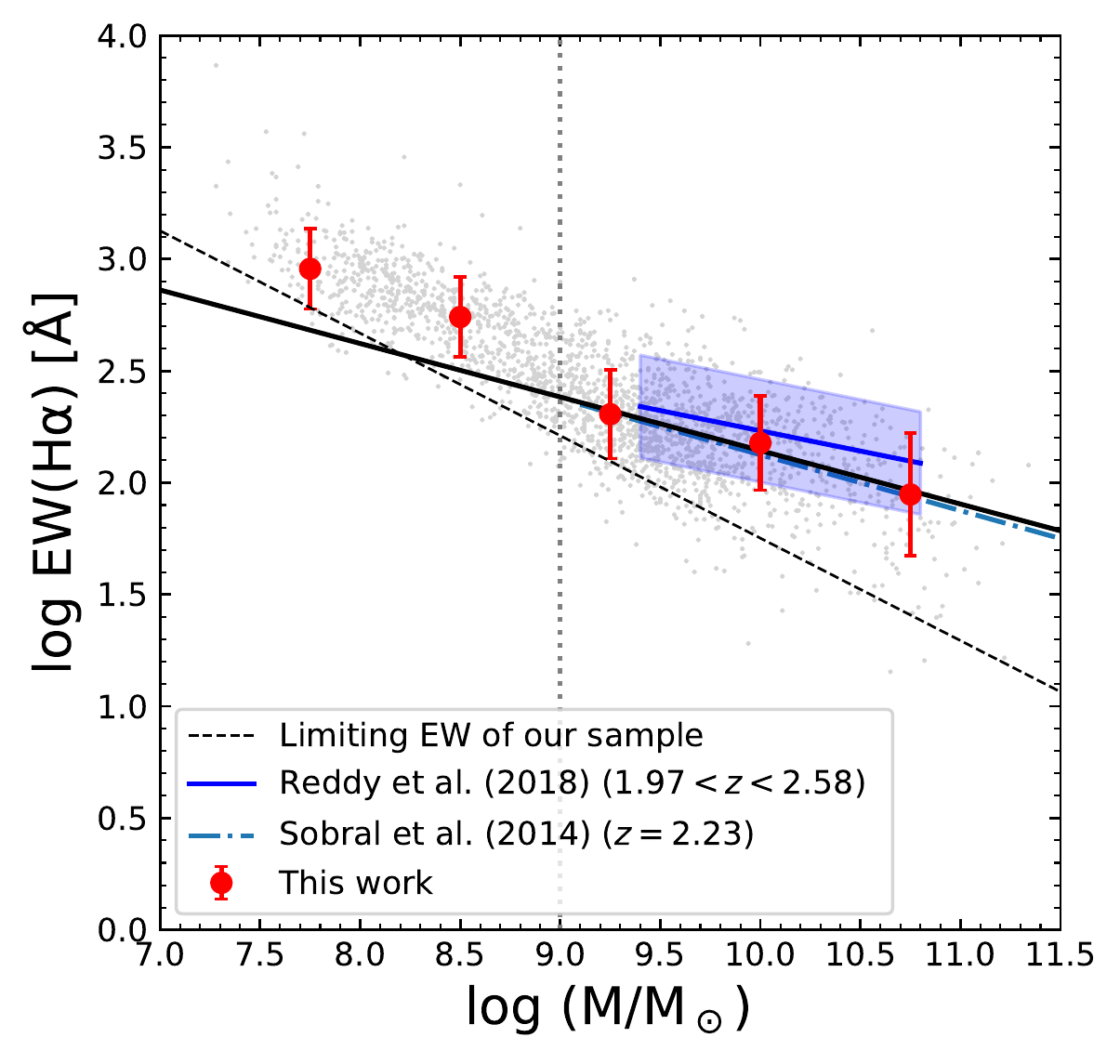}
  \caption{H$\alpha$ equivalent widths as a function of stellar mass. 
  Limiting equivalent widths of our sample is shown by the dashed black line as well.
  Since our selection is more sensitive to galaxies with strong continua, the limiting equivalent width depends on stellar mass.
  The grey and red circles represent individual galaxies and their medians in the mass bins, respectively.
  The best-fit relation obtained in the range of $\log{(M/M_\odot)}>9$ is shown by the black line extrapolated toward the low-mass end.
  The blue line and intervals show the typical H$\alpha$ equivalent width and its scatter reported by \citet{reddy_mosdef_2018} for galaxies with $\log{(M/M_\odot)}\simeq 9.3$--$10.7$.
  Our method seems sensitive enough to detect most of the H$\alpha$ emitters with $\log{(M/M_\odot)}>9.0$, but only extreme ones can be detected below this limit.}
  \label{fig:EWHa_vs_logM}
\end{figure}

Next, we estimate a limiting equivalent width above which the sample is complete.
As already mentioned, our equivalent width measurement is more sensitive to strong continuum (i.e., massive) galaxies.
Therefore the limiting equivalent width is expected to show stellar mass dependence.
By the definition of the emitter selection, a threshold of equivalent width to be selected as the emitter can be calculated as
\begin{eqnarray}
 EW_{\mathrm{lim}}(\mathrm{H\alpha}) = r_{\mathrm{H\alpha}} \times \frac{2\sigma_{K_{\mathrm{s}}}}{f_{\mathrm{cont}}} \times W / (1+z).
\end{eqnarray}
As the $K_\mathrm{s}$ photometry errors, {$\sigma_{K_\mathrm{s}}$}, do not vary significantly among the galaxies and $f_{\rm cont}$ correlates with stellar masses, $EW_{\mathrm{lim}}(\mathrm{H\alpha})$ is expected to show strong correlation with the stellar mass.
We have estimated $EW_{\mathrm{lim}}(\mathrm{H\alpha})$ for all the H$\alpha$ emitters as shown in \autoref{fig:EWHa_vs_logM}. In the data we find a correlation with stellar mass of the form:
\begin{eqnarray}
 \log{EW_{\mathrm{lim}} (\mathrm{H\alpha})} = -0.46 \log{(M/M_\odot)} + 6.33. 
\end{eqnarray}
\autoref{fig:EWHa_vs_logM} also shows our data generally agrees with estimates from the literature \citep{sobral_stellar_2014,reddy_mosdef_2018}.
Given this agreement, it seems our sample is at least mass complete at $\log{(M/M_\odot)}>9.0$, which corresponds to the 80\% mass completeness limit of ZFOURGE at $z\sim2$. A caveat on this conclusion is that we may still have missed emitters with relatively low equivalent widths.

In \autoref{fig:EWHa_vs_logM} we also fit to only galaxies with $\log{(M/M_\odot)}>9.0$ and find:
\begin{eqnarray}
 \log{EW (\mathrm{H\alpha})} = -0.24 \log{(M/M_\odot)} + 4.54.
\end{eqnarray}
This fit agrees very well with the results of previous works based on spectroscopy \citep{reddy_mosdef_2018} and narrow-band imaging \citep{sobral_stellar_2014}.

For galaxies with $\log{(M/M_\odot)}<9.0$, we cannot discuss their statistical properties due to mass incompleteness from the parent ZFOURGE sample.
Nevertheless it is interesting that there are many low-mass H$\alpha$ emitters with extremely high H$\alpha$ equivalent widths ($>500$ \AA). These values are much higher than the extrapolation of the relation at $\log{(M/M_\odot)}>9.0$ to lower masses.
Such galaxies are expected to have high sSFRs and we will further investigate them later.

\subsection{Attenuation by dust}
\label{sec:dust_HAE}
\begin{figure}
  \centering
  \includegraphics[width=0.5\textwidth]{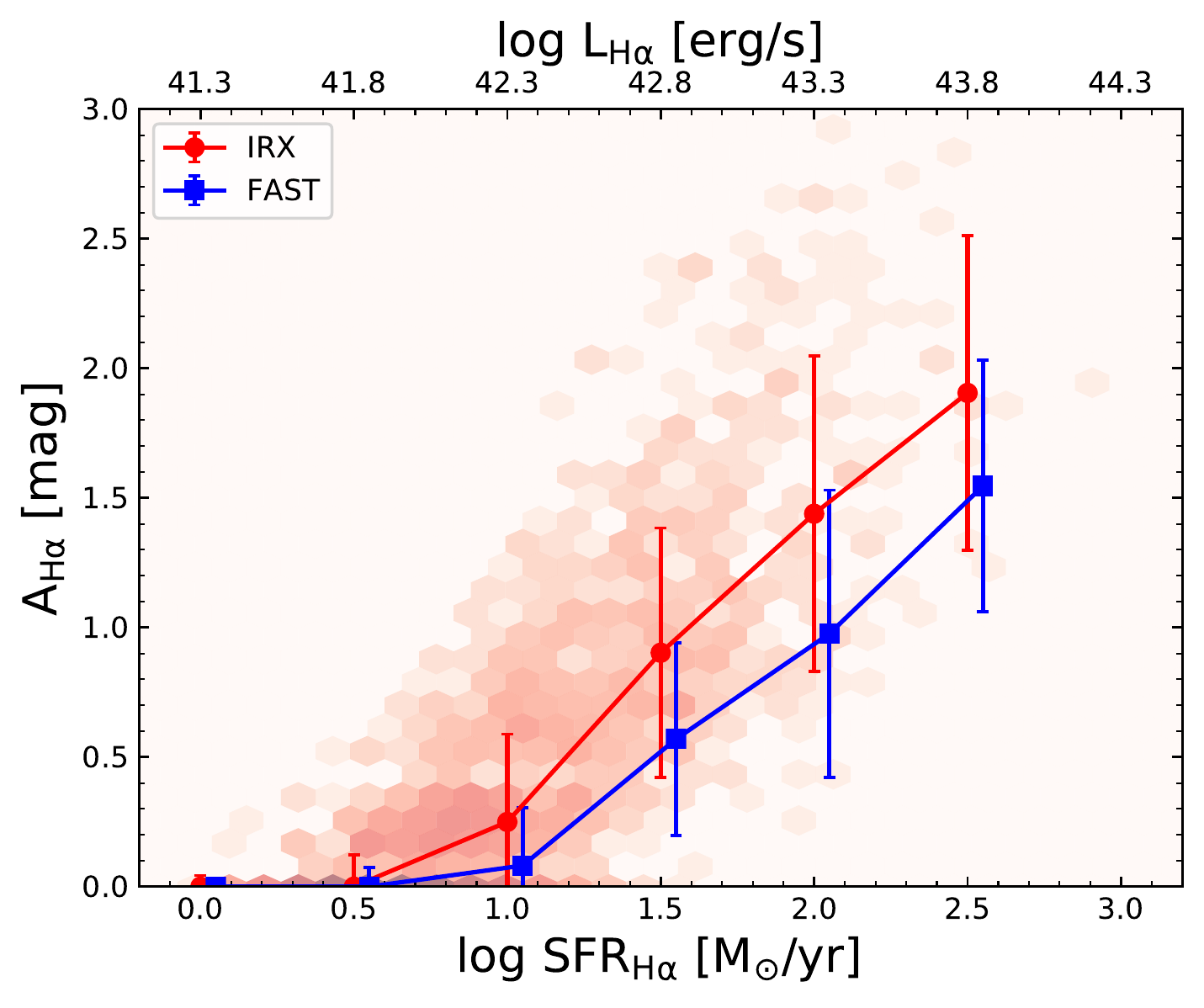}
  \caption{Correlation between H$\alpha$ SFRs and the amount of the dust attenuation estimated at the H$\alpha$ emission line, $A_{\mathrm{H\alpha}}$.
  The red hexagonal tiles show the $A_{\mathrm{H\alpha}}$ distribution of our sample, where $A_{\mathrm{H\alpha}}$ values are derived by the IRX as described in \ref{sec:dust_correction}.
  Median values of $A_{\mathrm{H\alpha}}$ in SFR bins derived from the IRX and the SED fitting by FAST are shown in the red circles and the blue squares, respectively.
  }
  \label{fig:sfrHa_AHa}
\end{figure}

\autoref{fig:sfrHa_AHa} shows a correlation between the H$\alpha$ SFR and estimated dust attenuation, which was derived two methods as described in Section \ref{sec:dust_correction}.
Clearly the galaxies with higher SFRs (i.e., more luminous in H$\alpha$) are more obscured.
This is unsurprising given that dust in the interstellar medium (ISM) of the galaxies is thought to be mainly produced by supernovae and AGB stars \citep[e.g.,][]{morgan_dust_2003}.
The best-fit relation between the H$\alpha$ SFRs and the amounts of the attenuation obtained from IRX relation in the sample is given by
\begin{eqnarray}
 A_{\mathrm{H\alpha}, {\mathrm IRX}} = 0.95 \log{\mathrm{SFR(H\alpha)}} - 0.51.
 \label{eq:sfrHa_AHa}
\end{eqnarray}
A similar relation is obtained for UV+IR SFRs using only the galaxies detected in the Spitzer/MIPS 24 $\mathrm{\mu m}$ band with S/N$>$3. 

The above relation is for the $A_{\mathrm{H\alpha}}$ derived from the IRX.
The difference is consistent with \autoref{fig:Av_relation}, where $A_{V, {\mathrm{FAST}}}$ shows systematically smaller values than $A_{V, {\mathrm{IRX}}}$.

\subsection{H\texorpdfstring{$\alpha$}{alpha} luminosity function}
\label{sec:LF_HAE}
Before discussing the luminosity function itself, we evaluate completeness of the sample with respect to H$\alpha$ luminosity by performing a simulation using observed relations.
First we generate 100,000 mock galaxies with stellar masses following a power-law distribution with the slope of $-$1.38 according to \citet{tomczak_galaxy_2014}.
Then we estimate their stellar continuum fluxes in the $K_\mathrm{s}$-band using a relation between the continuum fluxes and the stellar masses of our sample, which is obtained from the SED fitting carried out in Section \ref{sec:SED}.
Next, H$\alpha$ luminosities of the mock galaxies are drawn from a normal distribution with a mean and a standard deviation corresponding to the observed star formation main sequence (SFMS) and its scatter (see Section \ref{sec:SFMS}).
These luminosities are then converted to the excesses, where redshifts of the mock galaxies are randomly chosen from the observed distribution between 2.1 and 2.5.
Finally we obtain their total $K_\mathrm{s}$-band fluxes by adding the excesses to the stellar continuum fluxes.
We then choose galaxies which have (1) $K_\mathrm{s}$-band flux larger than 5$\sigma$, and (2) the excess fluxes larger than 1$\sigma$, where $\sigma$ is a photometric error, to be selected as H$\alpha$ emitters.
By calculating fractions of the emitters above the detection limit in each H$\alpha$ luminosity bin, we find that our sample is almost ($\sim 96$\%) complete for galaxies with $L_{\rm H\alpha} > 10^{42.25}\ \mathrm{erg\ s^{-1}}$, where the H$\alpha$ luminosities are corrected for attenuation by dust.

Number density of H$\alpha$ emitters in H$_{\alpha}$ luminosity is calculated by dividing the number of the emitters in each luminosity bin by the relevant comoving volume.
In our case, the total comoving volume, which is defined by the redshift range and the survey area, is $\sim 5.51 \times 10^5\ \mathrm{Mpc^3}$.
Then the number density in a bin is obtained as follows:
\begin{eqnarray}
\phi(\log(L_c)) = \frac{1}{\Delta(\log{L})} \sum_{|\log{\frac{L_i}{L_c}|} < \frac{\Delta \log{L}}{2}} \frac{1}{\Delta V},
\end{eqnarray}
where $L_c$ is a central value of the bin.
We count the number of emitters and divide it by the total volume ($\Delta V$) and the bin width
($\Delta \log{L} = 0.25$).
This formula is the same as the one used in \citet{sobral_hizels:_2009}.
Finally we fit a Schechter function to the number densities to determine the best-fit H$\alpha$ luminosity function, which is given as
\begin{equation}
\begin{split}
\phi(L)dL = \ln{10}\times \phi^* \left(\frac{L}{L^*}\right)^\alpha e^{-(L/L^*)} \left(\frac{L}{L^*}\right) d\log{L},
\end{split}
\end{equation}
where $L^*$, $\alpha$, and $\phi^*$ are a characteristic luminosity, a faint-end slope, and a normalization, respectively. 

\begin{figure}
  \centering
  \includegraphics[width=0.5\textwidth]{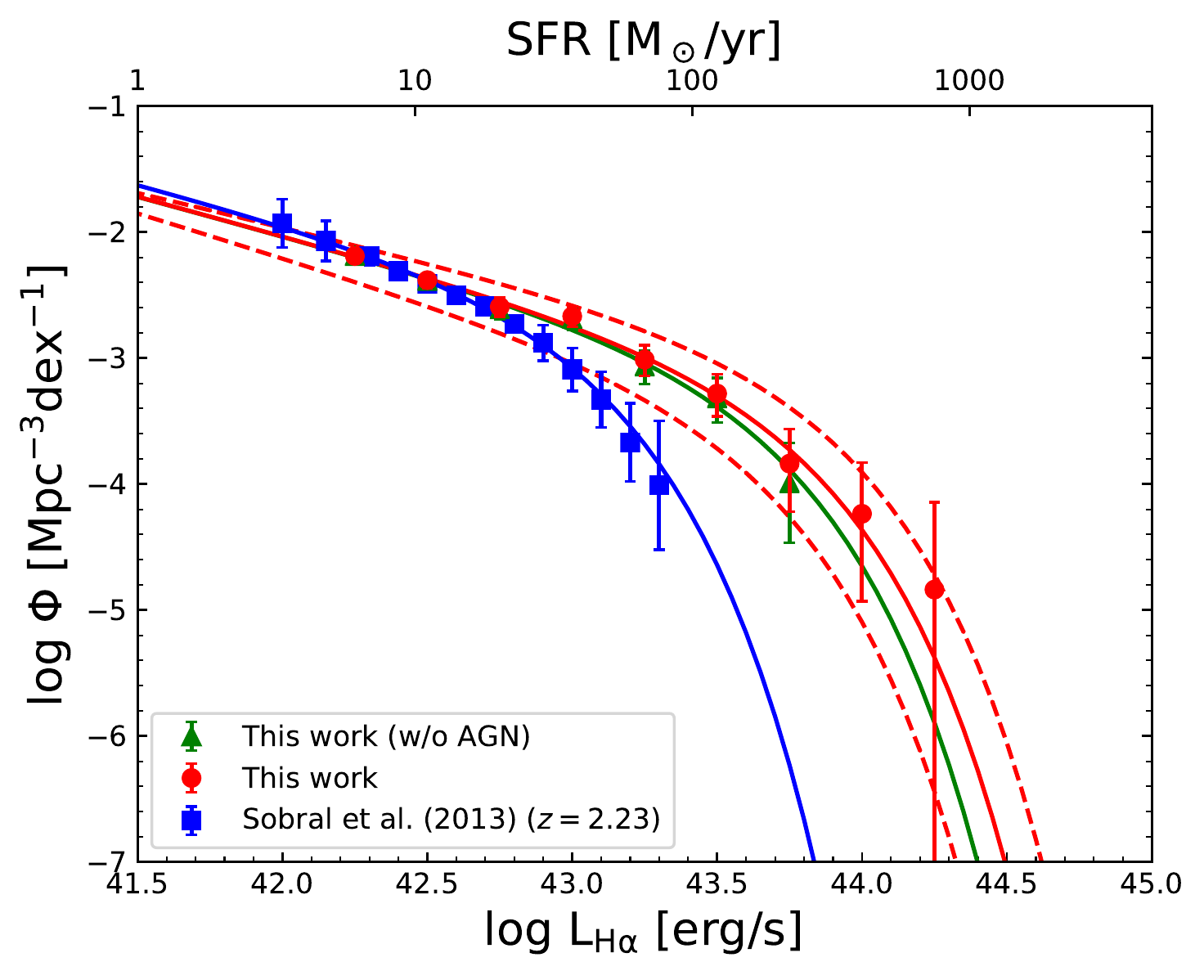}
  \caption{H$\alpha$ luminosity function of our sample with the result of \citet{sobral_large_2013} for a comparison, which are shown by red circles and blue squares, respectively.
  Green triangles illustrate how our result changes after removing AGNs as cataloged by \citet{cowley_zfourge_2016}.
  Each curve shows the best-fit Schechter function with the parameters summarized in \autoref{tab:LF_fit_param}, where 1$\sigma$ fitting uncertainties of our result are shown by red dashed curves.
  }
  \label{fig:LF_1_woAGN}
\end{figure}

\autoref{fig:LF_1_woAGN} shows our H$\alpha$ luminosity function with that of \citet{sobral_large_2013}.
The best-fit parameters are listed in \autoref{tab:LF_fit_param}.
\begin{table}[htb]
  \begin{center}
    \caption{Best-fit parameters of H$\alpha$ luminosity functions}
    \begin{tabular}{cccccccc}

        && $\alpha$ && $\log{L^*}$ && $\log{\phi^*}$ & \\ \hline \hline
        This work && -1.62 && 43.6 && -3.38 &  \\
        \ \\
        This work (w/o AGN) && -1.61 && 43.5 && -3.30 &  \\
        \ \\
        This work ($A_\mathrm{H\alpha}=1$) && -1.53 && 43.3 && -3.11 &  \\
        \ \\
        \citet{sobral_large_2013} && -1.59 && 42.9 && -2.78 &  \\
        \ \\
        \hline
    \end{tabular}
    \label{tab:LF_fit_param}
  \end{center}
\end{table}
It is immediately clear that our luminosity function has an excess in the bright-end compared to that of \citet{sobral_large_2013}, while the faint-end slopes are consistent.
We should note that the brightest two bins of our sample completely consist of AGNs as cataloged by \citet{cowley_zfourge_2016}.
Nevertheless, the excess still exists even if galaxies with known AGN are removed.

In the following we investigate inconsistency between our H$\alpha$ luminosity function and that of \citet{sobral_large_2013}.
We consider two possible causes.
One is related to structural properties of the galaxies, while the other is different corrections for dust extinction.

It has been proposed that galaxies form their stars in disks around central cores and increase their sizes, which has been called inside-out growth \citep[e.g.,][]{van_dokkum_growth_2010}.
At $0.7<z<1.5$, \citet{nelson_where_2016} have stacked imaging data of the 3D-HST and carried out S\'{e}rsic profile fitting 
to find radial profiles of H$\alpha$ emission lines to be more extended than stellar components.
Moreover, they have shown that the difference of the light profiles has become larger with increasing stellar mass.
Also at $z>2$, \citet{suzuki_extended_2019} have revealed extended H$\alpha$ profiles with AO-assisted imaging data.
These results indicate at a fixed aperture size one potentially might miss H$\alpha$ flux at large radii.

The flux measurements contained in the ZFOURGE catalog were originally performed with a $1.2''$ aperture on PSF-matched images and then corrected to total fluxes using growth curves from the $K_\mathrm{s}$-band \citep{straatman_fourstar_2016}.
In the HiZELS, \cite{sobral_hizels:_2009} used a fixed $2''$ aperture for all galaxies so they might have missed some fractions of the H$\alpha$ flux at radii $>2''$.
To estimate how much their fixed aperture can affect the luminosity function, we perform a simulation as follows.
First we assume that the stellar continuum profile of a galaxy follows a S\'{e}rsic profile \citep{graham_concise_2005}, that is
\begin{eqnarray}
I(R) = I_{\mathrm{e}} \left\{-b_n \left[ \left( \frac{R}{R_{\mathrm{e}}} -1 \right)^{1/n} \right]  \right\},
\end{eqnarray}
where $n$ is a S\'{e}rsic index, $R_\mathrm{e}$ a half-light radius, and $I_{\mathrm{e}}$ a surface intensity at that radius, $I(R_{\mathrm{e}})$.
Then $b_n$ is determined to satisfy the following equation:
\begin{eqnarray}
\Gamma(2n) = 2\gamma(2n,b_n),
\end{eqnarray}
where $\Gamma(n)$ and $\gamma(n,x)$ are complete and incomplete gamma functions, respectively.

\begin{figure}
  \centering
  \includegraphics[width=0.5\textwidth]{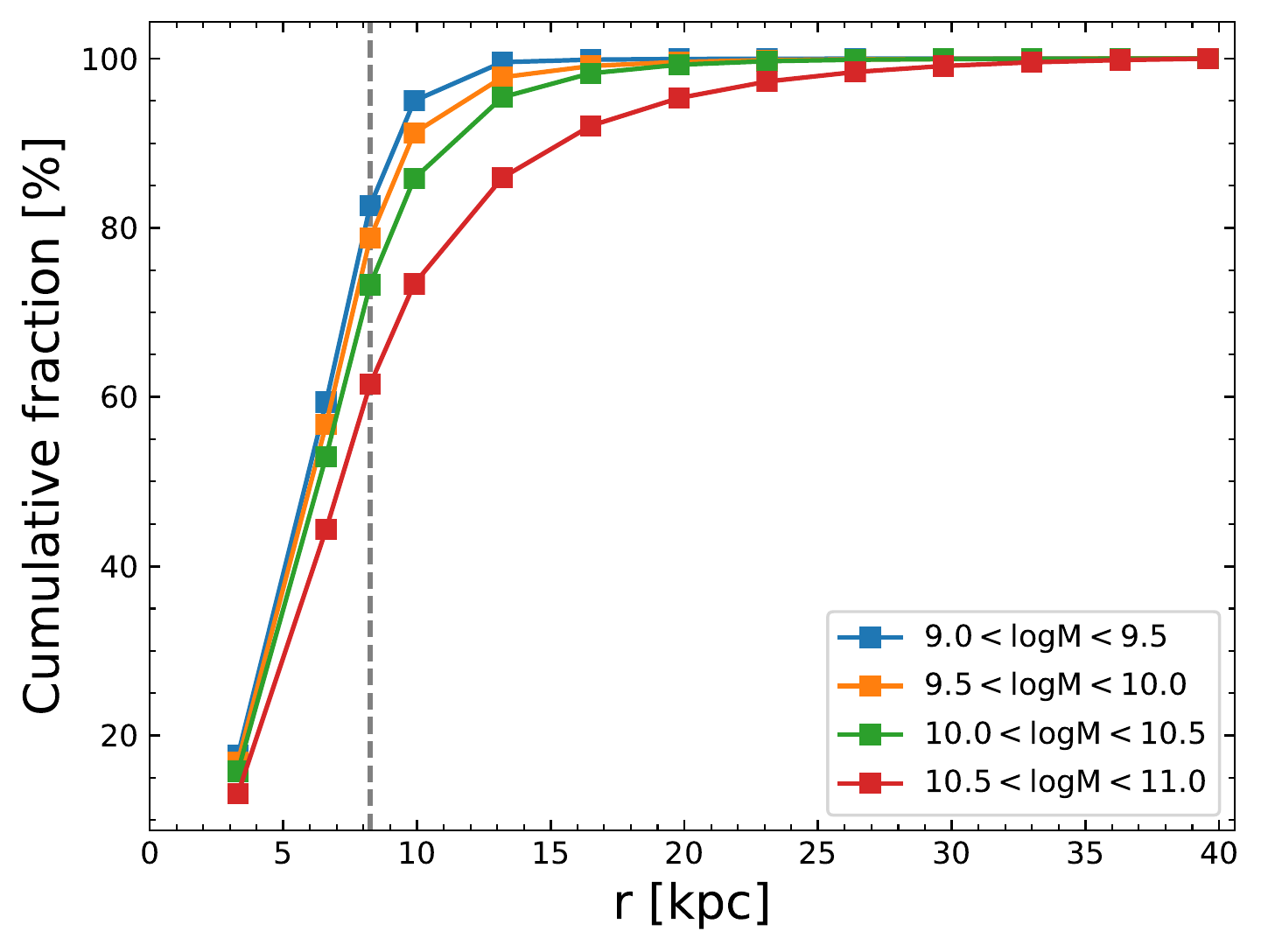}
  \caption{Cumulative fractions of H$\alpha$ fluxes as a function of radius in our simulation.
  They are obtained assuming the S\'{e}rsic profiles with the structural parameters of \citet{nelson_where_2016} and the seeing of the HiZELS observation.
  The color coding reflects stellar masses of the galaxies, while the grey vertical dashed line shows the physical scale corresponding to the $2''$ aperture used in \citet{sobral_large_2013} at a distance of $z=2.23$.
  }
  \label{fig:cumlative_Ha}
\end{figure}

We then create H$\alpha$ profiles having the S\'{e}rsic indices measured by \citet{nelson_where_2016}, which are average values derived from the stacked images of the 3D-HST.
To reproduce the observation, we convolve the profiles with a point spread function (PSF) corresponding to the seeing.
According to \citet{trujillo_effects_2001}, the seeing can be modeled as a moffat profile:
\begin{eqnarray}
\mathrm{PSF}(r) = \frac{\beta -1}{\pi^2 \alpha} \left[ 1+ \left(\frac{r}{\alpha} \right)^2 \right]^{-\beta},
\end{eqnarray}
where $\beta=2.5$.
A FWHM of the seeing is given using $\alpha$ and $\beta$ by
\begin{eqnarray}
\mathrm{FWHM} = 2\alpha \sqrt{2^{1/\beta} -1}.
\end{eqnarray}
In the case of the HiZELS, FWHM of the seeing is $\sim 1''$ \citep{geach_hizels:_2008}.
We create artificial images with the above models and perform photometry with varying aperture sizes.
As a result, we obtain cumulative fractions of H$\alpha$ fluxes as a function of radius shown in \autoref{fig:cumlative_Ha}.
A radius corresponding to the $2''$ aperture at $z=2.23$ is $\sim 8\ \mathrm{kpc}$ and represented by the grey dashed vertical line in \autoref{fig:cumlative_Ha}.
We can clearly see that at least $\sim 20$\% of the H$\alpha$ flux are missed by their aperture. 
This appears to be especially the case for the most massive galaxies, which are expected to reside in the bright-end of the luminosity function. We therefore conclude that it is possible that the fixed aperture used by \citet{sobral_large_2013} might have missed $\sim 40$\% of the H$\alpha$ flux.
We should note, however, that the measurements of radial profiles of the H$\alpha$ fluxes in \citet{nelson_where_2016} are truncated at a radii of $\sim 10\ \mathrm{kpc}$, and we have extrapolated their profiles up to 40 kpc for our simulation.

\begin{figure}
  \centering
  \includegraphics[width=0.5\textwidth]{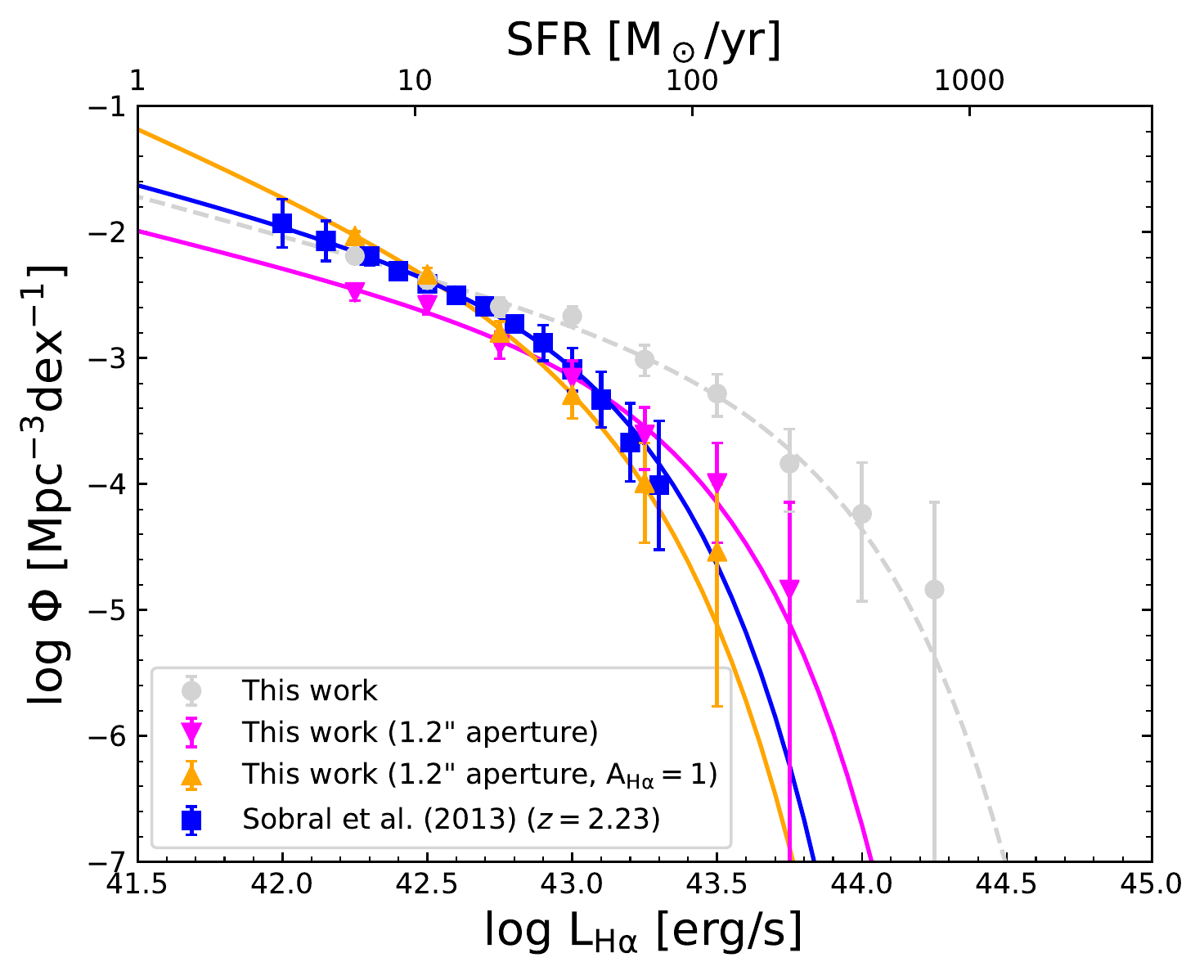}
  \caption{H$\alpha$ luminosity function derived from the ZFOURGE data without the aperture corrections assuming $A_{\mathrm{H\alpha}}=1$ (orange triangles).
  The difference between this work (grey circles) and \citet{sobral_large_2013} (blue squares) is almost completely gone.
  }
  \label{fig:LF_fixed}
\end{figure}

The effects of the missing fluxes can be also clearly seen in \autoref{fig:LF_fixed}, where we show a luminosity function derived from ZFOURGE data without the aperture corrections (magenta triangles).
The luminosity function is suppressed as a whole due to missing H$\alpha$ flux.

A potential caveat of this analysis is the assumption we make in about the H$\alpha$ sizes of our $z=$2.1--2.5 sample.
It is well-established that galaxies evolve in size at least in terms of their continuum properties \citep[e.g.,][]{van_der_Wel_2014}.
As we are using sizes from a lower redshift from \citet{nelson_where_2016}, there is a chance that at $z\sim2$ these average sizes are not appropriate, which will have implications for the analysis of missing H$\alpha$ flux.
Indeed, \citet{Wilman_2020} have found H$\alpha$ seems to follow continuum sizes, which means our H$\alpha$ sizes we adopt for this analysis are likely overestimated.
This would act to reduce the amount of missing H$\alpha$ flux in the fixed size aperture.
However, from \citet{van_der_Wel_2014}, we estimate the size evolution for star-forming galaxies over $z=1.5$ to $z=2$ only changes by $\sim10\%$, which means more than 20\% of H$\alpha$ flux could have been missed even at $z>2$.
For comprehensive understanding on the missing flux in H$\alpha$, more observations are necessary to reveal structural properties of galaxies at $z>2$.

Another difference between this work and \citet{sobral_large_2013} is the dust extinction correction.
As shown in \autoref{fig:sfrHa_AHa} there is a positive correlation between H$\alpha$ luminosity and the estimated attenuation by dust.
\citet{sobral_large_2013} adopted $A_{\mathrm{H\alpha}}=1\ \mathrm{mag}$ for all the galaxies, which is not supported by our data.
They therefore have likely underestimated intrinsic luminosities of the brightest H$\alpha$ emitters and overestimated the luminosities for the faintest galaxies. To demonstrate how the adoption of a fixed attenuation impacts a luminosity function, we show a luminosity function derived from ZFOURGE data without the aperture corrections and assuming a fixed H$\alpha$ attenuation of 1 mag with orange points in \autoref{fig:LF_fixed}. This modified luminosity function agrees with that of HiZELS very well.

From the above results, we conclude that the bright-end discrepancy between our luminosity function and that derived in \citet{sobral_large_2013} is caused by intrinsically extended H$\alpha$ profiles and the larger $A_{\mathrm{H\alpha}}$ in bright H$\alpha$ emitters.
This implies there are larger numbers of intrinsically bright H$\alpha$ emitters than previously reported.
The main caveat on this conclusion is that number of such objects is small, therefore the impact of cosmic variance and AGNs might complicate our interpretation.
Larger surveys and spectroscopic follow-ups are required to better understand the population in the bright-end.

\subsection{Cosmic star formation rate density}
By integrating the luminosity function, 
we can derive a cosmic SFRD at $2.1<z<2.5$.
Here we remove a contribution from identified AGNs, which is $\sim 13$\% of the total H$\alpha$ luminosity.
In addition, we perform the integration with two intervals because the faint-end slope of the luminosity function is uncertain.
One is the range of $L_{\mathrm{H\alpha}}=0$--$10^{45}\ \mathrm{erg\ s^{-1}}$, while the other is of $L_{\mathrm{H\alpha}}=10^{41.6}$--$10^{45}\ \mathrm{erg\ s^{-1}}$.
These intervals are same as those used in \citet{sobral_large_2013}, where $L_{\mathrm{H\alpha}}=10^{41.6}\ \mathrm{erg\ s^{-1}}$ corresponds to $0.01L^*$.

\begin{figure}
  \centering
  \includegraphics[width=0.5\textwidth]{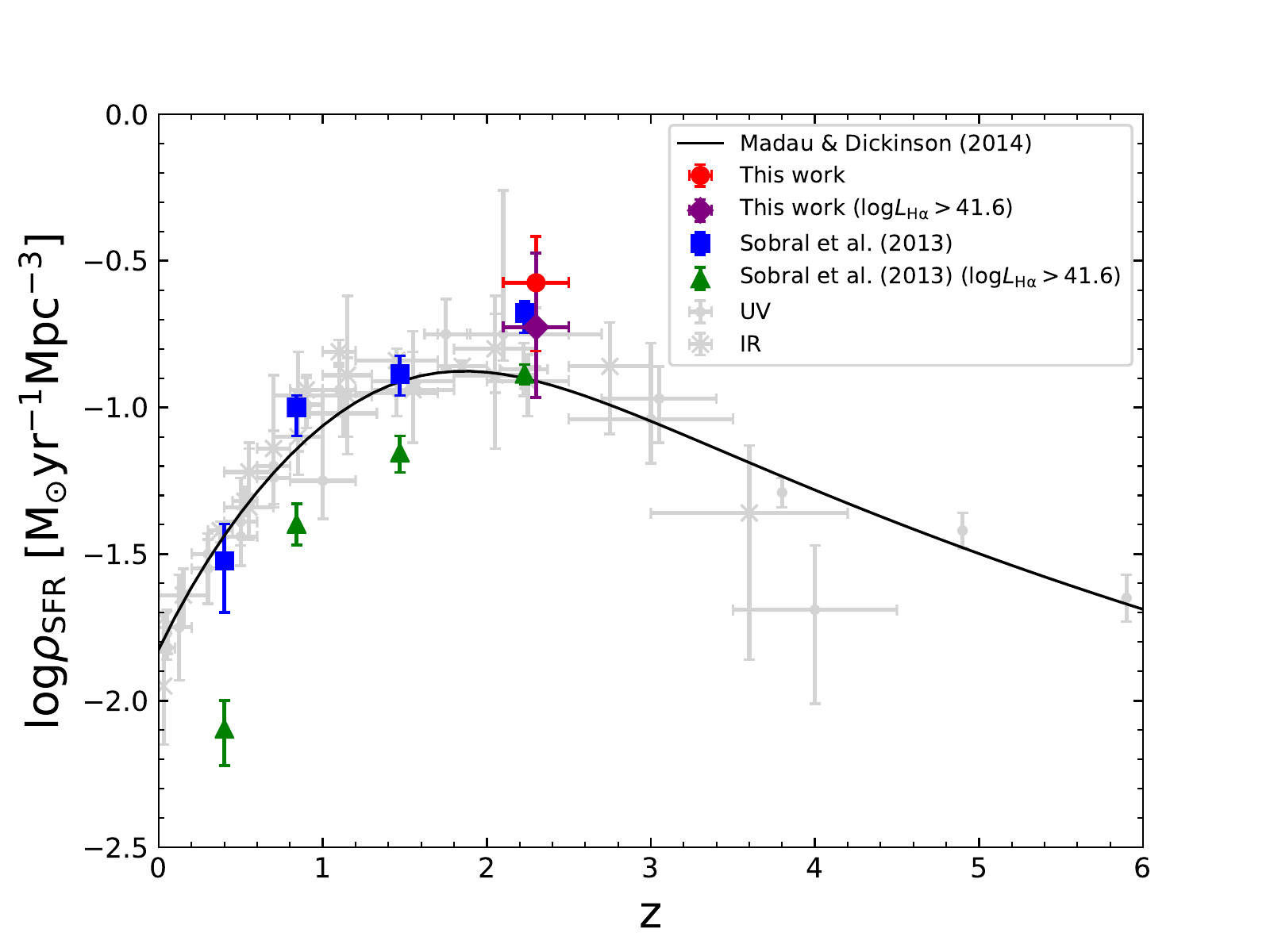}
  \caption{Our cosmic star formation rate density measurement compared with previous studies.
  Red and magenta circles represent our measurement with integration range of $L_{\rm H\alpha}=0-10^{45}$ (erg s$^{-1}$) and $10^{41.6}-10^{45}$ (erg s$^{-1}$), respectively.
  Blue squares and green triangles represent the results of \citet{sobral_large_2013}, with same integration range as ours.
  AGN contributions have been removed from both values.
  The grey circles and crosses show the results of previous studies summarized in \citet{madau_cosmic_2014} for a comparison, which are measured from UV and/or IR luminosities.
  }
  \label{fig:sfrd_Ha}
\end{figure}

Given that the H$\alpha$ emission line is a direct SFR tracer and that H$\alpha$ emitters cover diverse populations of star-forming galaxies \citep{oteo_nature_2015}, we expect a consistent H$\alpha$ SFRD with previous studies based on UV and/or IR observations.
\autoref{fig:sfrd_Ha} shows our SFRD with those previously reported as summarized in \citet{madau_cosmic_2014}.
We find that our estimate agrees well with previous work, suggesting that our H$\alpha$ emitter selection reflects the underlying star-forming galaxy population and that our correction for the dust extinction properly works to recover total H$\alpha$ SFRs.
However, we cannot rule out the possibility that galaxies strongly obscured by the dust are missing from our sample because our selection requires galaxies to be detected in the $K_\mathrm{s}$-band.

The SFRD agreement is interesting given we find a larger number of bright H$\alpha$ emitters than previous work. This reflects the relatively small contribution of the brightest galaxies to the total star formation activity due to their  small numbers.

\subsection{Star formation main sequence}
\label{sec:SFMS}
We also investigate the relation between stellar mass and star formation rate of our H$\alpha$ emitter sample.
In the following analysis, all the AGNs identified by either X-ray, IR, or radio (38/2005) are removed while the quiescent H$\alpha$ emitters are included.

\begin{figure}
  \centering
  \includegraphics[width=0.5\textwidth]{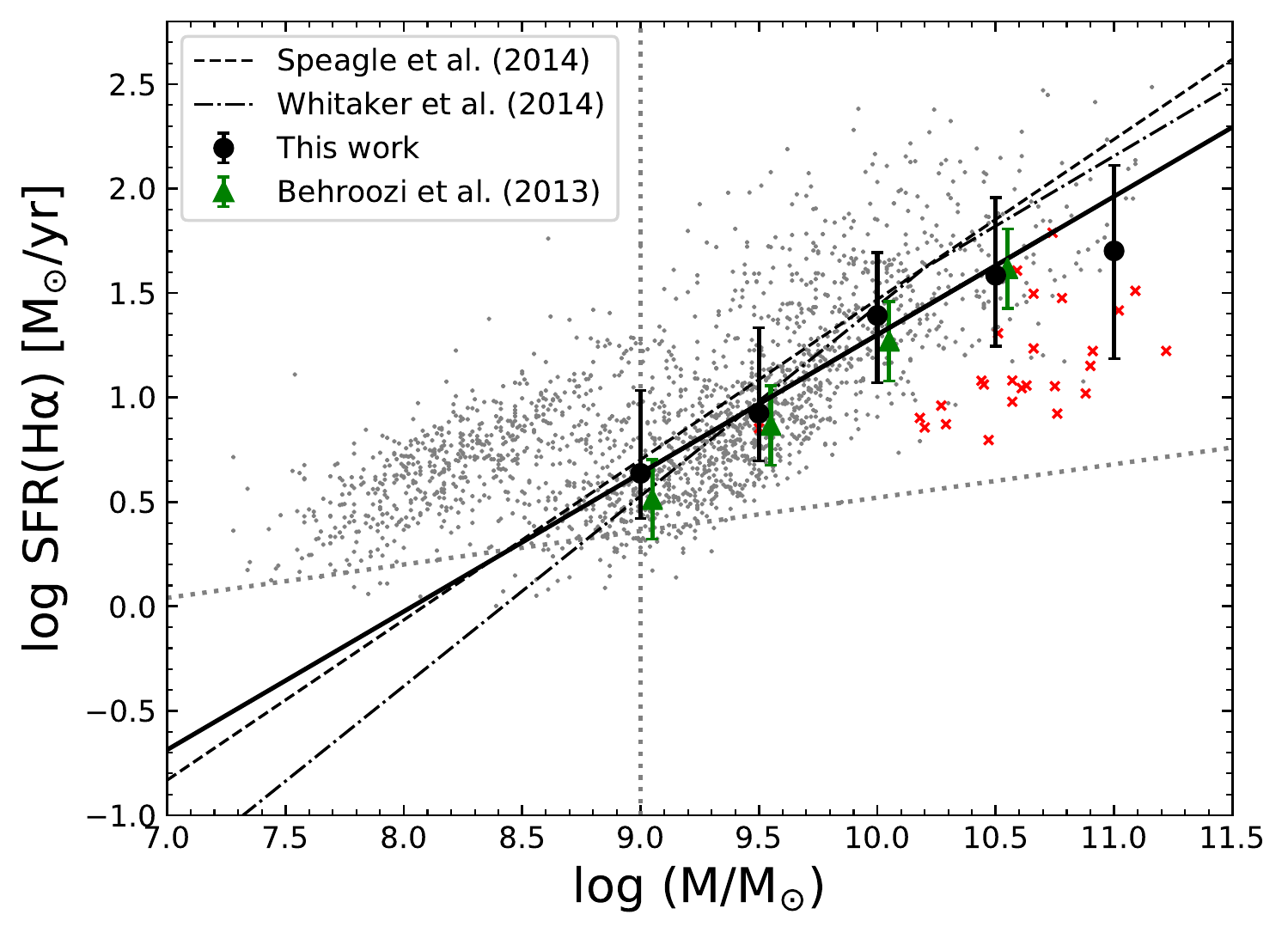}
  \caption{H$\alpha$ star formation main sequence (SFMS) of our sample.
  Grey dots represent individual galaxies while black circles show median values in each stellar mass bin.
  By fitting in the range of $9<\log{(M/M_\odot)}<11$, we obtain the best-fit SFMS shown by the black solid line. 
  Green triangles are the typical values among many previous studies using observations at various wavelengths \citep{behroozi_average_2013}.
  Black dashed and dash-dotted lines indicate the best-fit SFMSs of \citet{speagle_highly_2014} and \cite{whitaker_constraining_2014}, respectively.
  Red crosses represent the quiescent H$\alpha$ emitters with well-constrained SEDs classified by the $UVJ$ diagram.
  Dotted vertical line shows the stellar mass completeness limit of our sample while dotted horizontal line shows the SFR limit of our sample. This limit corresponding to the flux excess threshold to be selected as a H$\alpha$ emitter. 
  }
  \label{fig:SFMS_Ha}
\end{figure}

\autoref{fig:SFMS_Ha} shows a stellar masses--H$\alpha$ SFRs plot of our sample with SFMS of previous studies \citep{speagle_highly_2014,whitaker_constraining_2014,behroozi_average_2013}.
The black line represents our best-fit SFMS, which is given by
\begin{eqnarray}
    \label{eq:SFMS}
    \log{\mathrm{SFR(H\alpha)}} = 0.66 \log{(M/M_\odot)} - 5.32.
\end{eqnarray}
In the fitting, we only used the galaxies in the range of $9< \log{(M/M_\odot)} < 11$ because the incompleteness becomes significant below this range. At $\log{(M/M_\odot)} > 11$ the number of the galaxies is small and a large fraction of them are classified as quiescent.
Our result is consistent with the previous studies.
Especially, a good agreement with \citet{whitaker_constraining_2014}, where SFRs were derived from UV+IR luminosities, indicates that our correction for the dust extinction successfully recovers total H$\alpha$ SFRs.
The small offset in the most massive bins, though within the scatter, is probably due to different selections of the star-forming galaxies.
For example, selection by observed FIR fluxes may bias a sample toward dusty starburst systems rather than normal galaxies on the SFMS \citep{rodighiero_lesser_2011,lee_multi-wavelength_2013}.
In contrast, H$\alpha$ selection is thought to reflect a more diverse galaxy population \citep{oteo_nature_2015}.

\subsection{Low-mass galaxies with high sSFR}
At $\log{(M/M_\odot)} > 9$, we find that the scatter around the SFMS is $\sim 0.3\ \mathrm{dex}$, which is defined as a standard deviation of SFR offsets of individual galaxies from the SFMS.
This value agrees well with previous observations \citep[e.g.][]{Shivaei_2015} and support a ``self-regulated'' evolution as outlined in \citet{tacchella_confinement_2016}.

At $\log{(M/M_\odot)}<9$, however, there are many low-mass galaxies with much higher SFRs than predicted from an extrapolation of the SFMS.
We should note, of course, that only strong H$\alpha$ emitters can be detected in the low-mass end due to the sensitivity limit of the $K_\mathrm{s}$-band of ZFOURGE.
Nevertheless it is still interesting to investigate them further because the high sSFRs may suggest their bursty star formation histories (SFHs).
Several works have also favor a bursty SFH interpretation for low-mass galaxies both observationally and theoretically \citep[e.g.][]{Broussard_2019, faisst_recent_2019, emami_closer_2019, sparre_starbursts_2017}.

\begin{figure}
  \centering
  \includegraphics[width=0.5\textwidth]{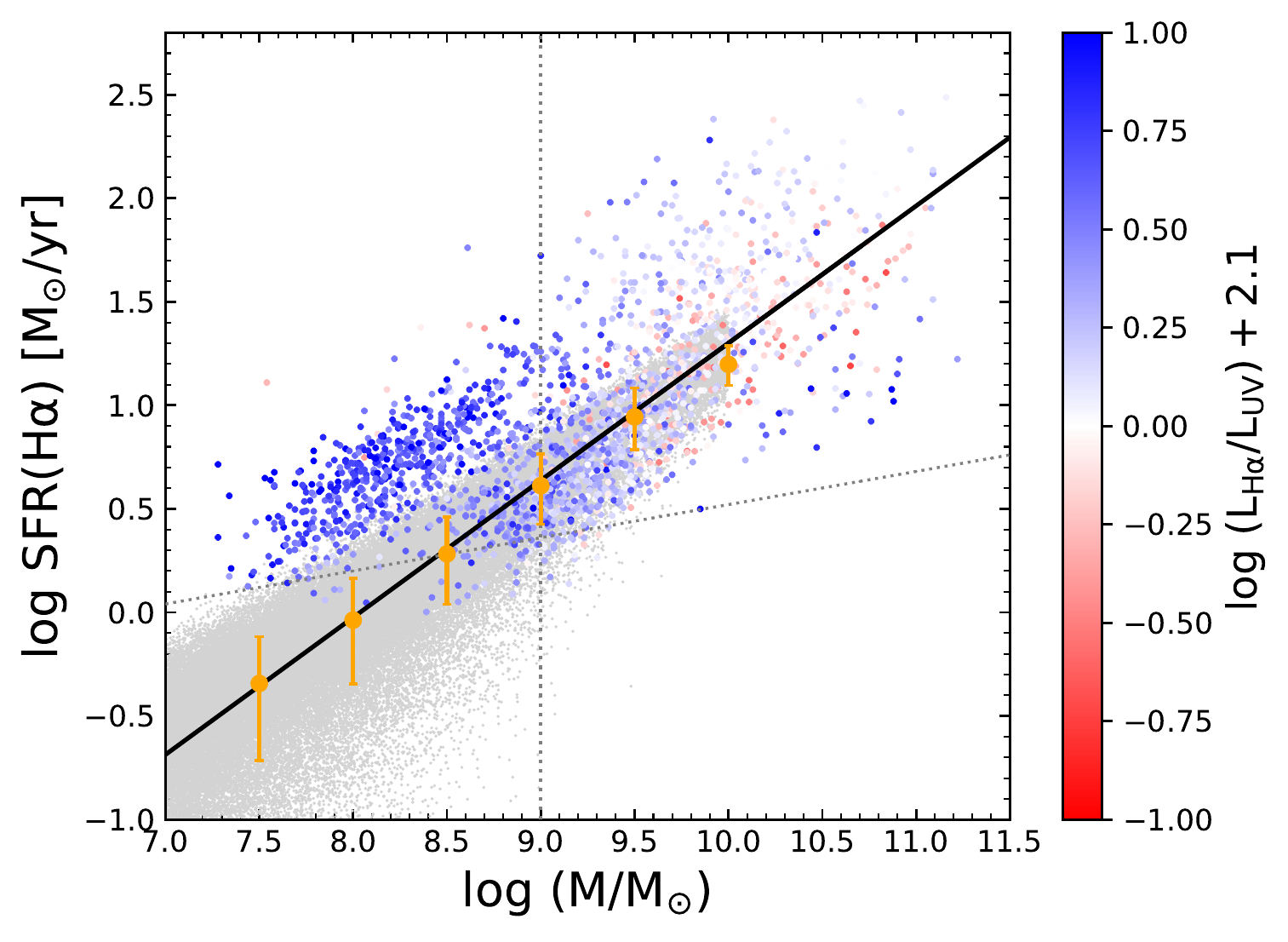}
  \caption{SFMS plot same as \autoref{fig:SFMS_Ha} but individual observed galaxies are colored according to luminosity ratios between H$\alpha$ and UV (see Section \ref{sec:ha2uv}). Simulated observations are shown as grey dots with the scatter in certain stellar mass bins shown as orange bars. The difference between the simulations and observations at $\log{(M/M_\odot)}<9$ suggest that high photometric scatter is not the likely causing the high SFRs at low masses. 
  }
  \label{fig:SFMS_simulation}
\end{figure}

Before investigating the SFHs, we first evaluate the impact of photometric errors on the sSFRs of the low-mass galaxies.
It might be possible that the apparently high sSFRs are only caused by increased photometric errors in their H$\alpha$ fluxes.
To explore this possibility, we perform a simulation as follows.
First we generate 100,000 mock galaxies with stellar masses of $\log{(M/M_\odot)}=7.0$--$10$ following the stellar mass function of \citet{tomczak_galaxy_2014}.
Their SFRs are calculated by \autoref{eq:SFMS}, where we assume that our best-fit linear relation to the SFMS is held even at $\log{(M/M_\odot)}<9$.
Then we add fluctuations to the simulated SFRs based on the photometric errors in the $K_{\mathrm{s}}$-band (grey circles in \autoref{fig:SFMS_simulation}).
Next we derive probability distributions from the intrinsic scatter of the simulated observation in different stellar mass bins (orange bars in \autoref{fig:SFMS_simulation}). 
The larger observed scatter ($\gtrsim 0.5\ \mathrm{dex}$) compared to the simulated scatter from photometric errors at $\log{(M/M_\odot)}<9.0$ likely means the high SFR values are not due to photometric errors.
Therefore we conclude that the low-mass galaxies with the high sSFRs show starburst-like sSFRs.

\subsection{H\texorpdfstring{$\alpha$}{alpha}/UV ratio}
\label{sec:ha2uv}
If a galaxy has undergone an instantaneous starburst within the past 100 $\mathrm{Myr}$, its luminosity ratio between H$\alpha$ and UV is expected to be different from a galaxy that is forming stars at a constant rate \citep{sparre_starbursts_2017,faisst_recent_2019}.
This is because an instantaneous burst is reflected effectively immediately in the H$\alpha$ luminosity compared to the observed UV.
As a consequence, when a short starburst occurs, the H$\alpha$/UV ratio is boosted (relative to a constant SFH) during first 10 $\mathrm{Myr}$. Subsequently the ratio is returns to a lower value 10--100 $\mathrm{Myr}$ after the burst.
This means that if the high sSFRs of the low-mass galaxies have been caused by recent starbursts, they should also have high H$\alpha$/UV ratios.

\begin{figure}
  \centering
  \includegraphics[width=0.5\textwidth]{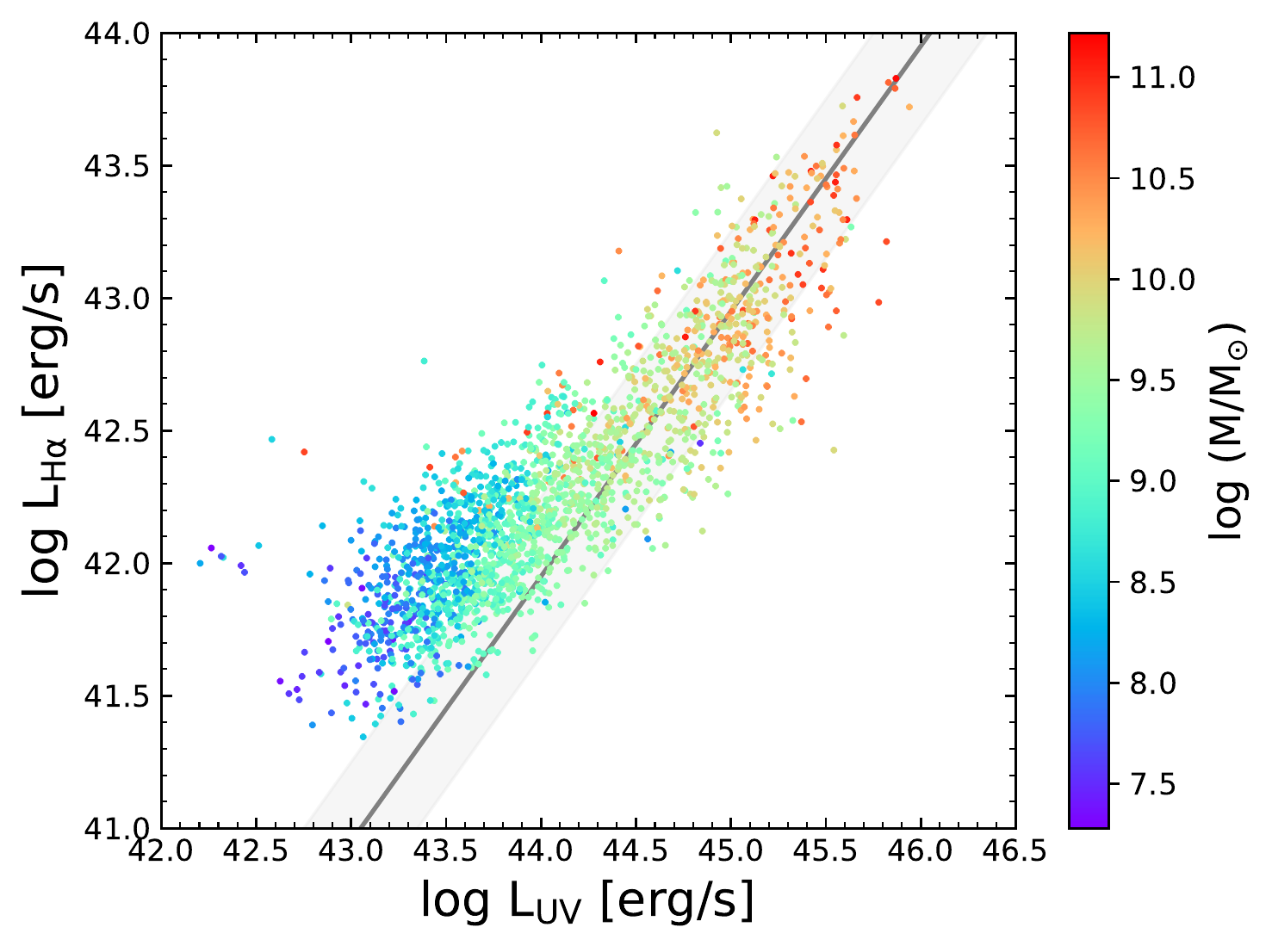}
  \caption{Comparison between the H$\alpha$ and UV luminosities, which are both corrected for the dust attenuation.
  The color coding reflects stellar masses of the galaxies.
  The grey line shows the expected relation from a constant SFH without starbursts within $100\ \mathrm{Myr}$ \citep{faisst_recent_2019}. The shaded region indicates a $\pm 0.3\ \mathrm{dex}$ region about the relation.
  }
  \label{fig:LHa_vs_LUV}
\end{figure}

\autoref{fig:LHa_vs_LUV} shows a comparison of the H$\alpha$ luminosities and the UV luminosities of our H$\alpha$ emitters, where both have been corrected for the dust extinction.
Grey line represents the expected relation when assuming a constant SFH \citep{faisst_recent_2019} and shaded area shows a $\pm 0.3$ dex region about the relation.
Most of the galaxies with $\log{(M/M_\odot)}<9.0$ have very high H$\alpha$/UV ratios, which can not be explained by a constant SFH. Very few massive galaxies show such high ratios.
\autoref{fig:SFMS_simulation} makes it clear that the low-mass galaxies that are above the extrapolated SFMS correspond to the population with high H$\alpha$/UV ratios.
These results suggest that the low-mass H$\alpha$ emitters with high sSFRs found here likely recently experienced short period starbursts.

\subsection{SFHs of the starburst galaxies}
Here we bring the results together and explore the possibility that there are two starbust modes: one for low-mass galaxies and another for high-mass galaxies. While in the previous section we mostly focused on low-mass starburst, there are also high-mass galaxies with high sSFRs. From \autoref{fig:SFMS_simulation} it is clear their H$\alpha$/UV ratios are lower than those observed for the low mass galaxy starbursts.
This may imply that starburst galaxies with different stellar masses have different SFHs.
In other words, the mechanisms which triggered the starbursts may depend on the stellar masses of the galaxy.

\begin{figure}
  \centering
  \includegraphics[width=0.5\textwidth]{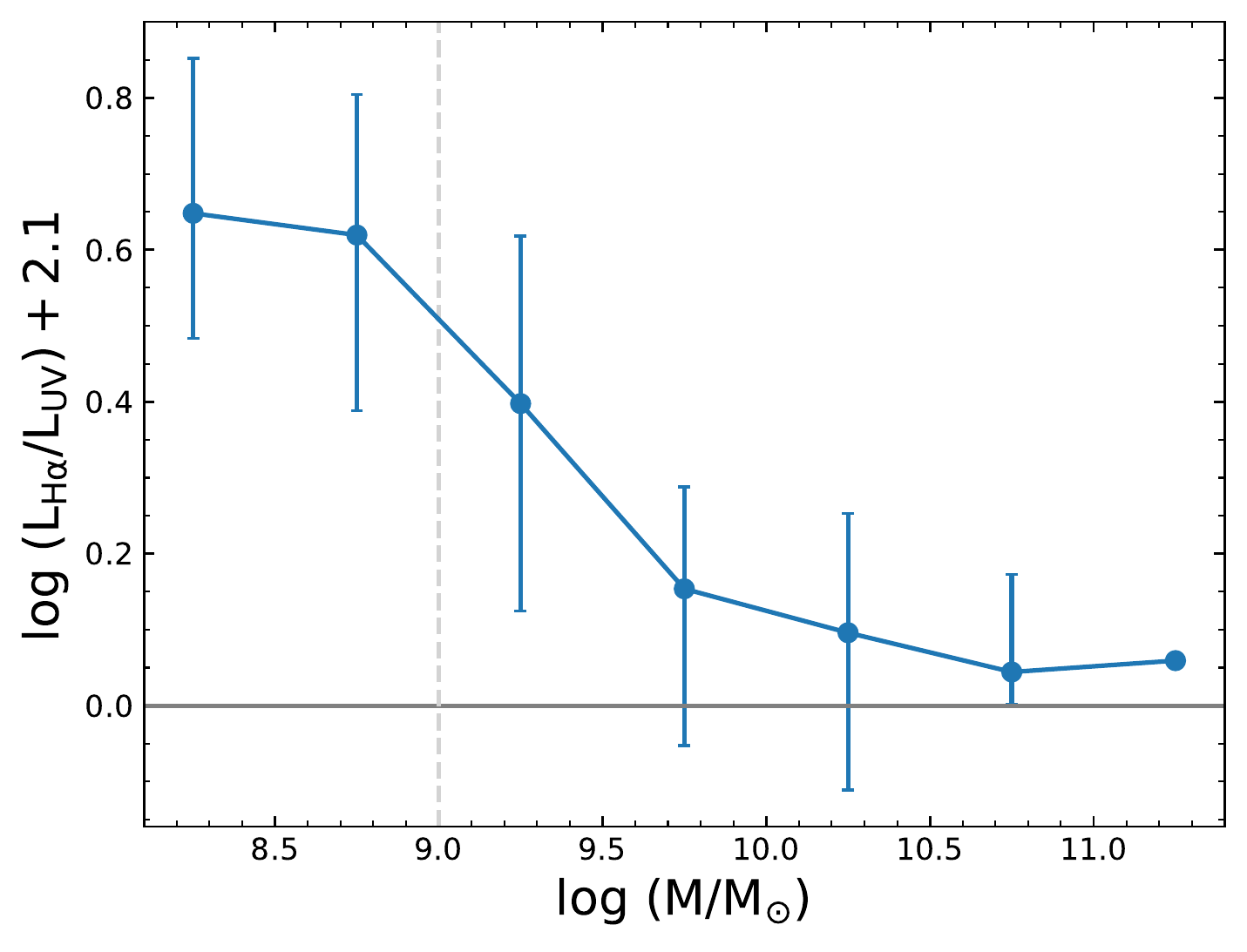}
  \caption{H$\alpha$/UV ratios of the starburst galaxies as a function of stellar mass.
  The blue circles represent median values in each stellar mass bin.
  The grey horizontal line shows the value expected from a constant SFH, while the dashed vertical line indicates the stellar mass completeness limit.
  }
  \label{fig:Lratio_starburst}
\end{figure}

\begin{figure*}[ht!]
  \centering
  \includegraphics[width=\textwidth]{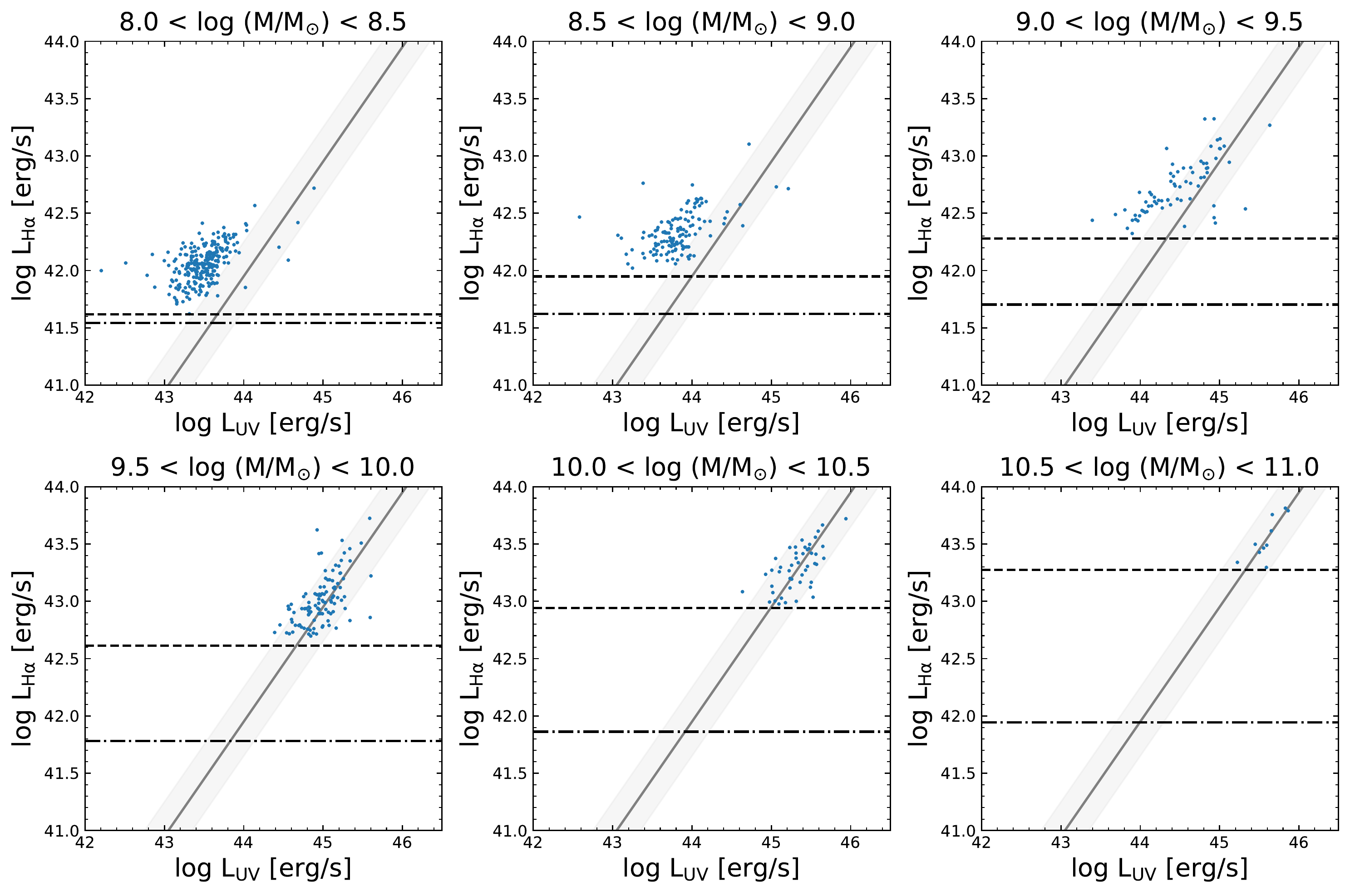}
  \caption{Same as \autoref{fig:LHa_vs_LUV}, but only the starburst galaxies are plotted separately according to their stellar masses.
  Dashed black horizontal lines represent thresholds of the H$\alpha$ SFR to be classified as the starburst galaxies in our definition while dashed-dotted horizontal lines is our H$\alpha$ emitter selection. As the luminosity criteria used for starburst selection are always higher than the limiting luminosity for HAE selection, we conclude the ratios in \autoref{fig:Lratio_starburst} are not likely due to a selection bias.}
  \label{fig:Lratio_bins}
\end{figure*}

To investigate this idea more quantitatively, we define the starburst galaxies as those with SFRs larger than expected from the SFMS by 0.3 dex and measure the median H$\alpha$/UV ratio in each stellar mass bin, as shown in \autoref{fig:Lratio_starburst}.
It is clear that H$\alpha$/UV decreases with increasing stellar masses and approaches a value expected from a constant SFH. Before interpreting this trend, it is important to understand if any selection affects the high H$\alpha$/UV ratios of the starburst galaxies, as H$\alpha$/UV ratios can be overestimated if UV-faint galaxies at fixed H$\alpha$ luminosities and/or H$\alpha$-bright galaxies at fixed UV luminosities are preferentially selected somehow.
The former is unlikely because the optical data
are much deeper than the IR, so we do not explore this possibility here.
The latter, a preferential selection of H$\alpha$-bright galaxies at fixed UV luminosities is plausible, so we explore it further in \autoref{fig:Lratio_bins}. 
The Figure shows the H$\alpha$/UV ratios of the starburst galaxies in various stellar mass bins.
As the starburst galaxies are, by definition, selected to have high H$\alpha$ luminosities (i.e. SFRs), the luminosity criteria used for starburst selection are always higher than the limiting luminosity for HAE selection.
Given this finding, we conclude that any selection bias inherent to the sample should have no significant impact on the H$\alpha$/UV ratios.
Therefore the impact of the bias toward luminous H$\alpha$ emitters on the H$\alpha$/UV ratios seems insignificant. We therefore conclude that the trend in \autoref{fig:Lratio_starburst} is real.

The results explored here seem to imply there are two types of starburst modes. As mentioned above, a high H$\alpha$/UV ratio implies a bursty SFH, suggesting that a starburst has occurred within past $\sim 10\ \mathrm{Myr}$.
Therefore the trend shown in \autoref{fig:Lratio_starburst} indicates that there are at least two different types of starbursts.
One is a burst with a short duration, which is dominant in low-mass galaxies, causing high H$\alpha$/UV ratios.
The other is a burst with a longer duration, which is dominant in high-mass galaxies, resulting in constant SFHs lasting for $> 10\ \mathrm{Myr}$.
Although it is beyond the scope of this work to reveal specific mechanisms that govern starburst activities in each stellar mass bin, several phenomena have been proposed as triggers such as supernova feedback, gas inflows, and major mergers.

%% file: summary_astroph.tex
\section{Summary}
\label{sec:summary}

We have identified 2005 H$\alpha$ emitters at $2.1<z<2.5$ from the ZFOURGE catalog. We take advantage of the deep ZFOURGE $K_{\mathrm{s}}$ photometry and wide wavelength coverage and derive 
H$\alpha$ flux estimates from observed $K_{\mathrm{s}}$ photometry. We specifically measure the $K_{\mathrm{s}}$ flux excess which is the difference between observed fluxes and a stellar continuum flux estimated by SED fitting.
When performing the SED fitting, contributions of emission lines were included in the SED templates.

We have corrected H$\alpha$ luminosity for dust extinction based on the IRX method.
For galaxies without IR detection, we used a relation between $A_V$ derived from the IRX and the SED fitting.
As a result, we have found that the attenuation of H$\alpha$ depends on both H$\alpha$ luminosity and UV+IR SFR, ranging from 0 to 2 mag.
The correction for dust extinction is more important for galaxies with higher SFR.

The H$\alpha$ luminosity function shows an excess in the luminous end compared to the result of \citet{sobral_large_2013}.
The excess is mostly explained by missing H$\alpha$ flux in \cite{sobral_large_2013} due to their use of a fixed photometric aperture and their not accounting for a SFR dependence on the strength of the dust extinction.
Specifically, our simulation suggests that the $1.2''$ aperture used in \citet{sobral_large_2013} would have missed $\sim 40$\% of the H$\alpha$ fluxes for the luminous-end galaxies.
Moreover, they have underestimated intrinsic H$\alpha$ luminosities of the most luminous galaxies by at most 1 mag as they assumed $A_{\mathrm{H\alpha}} = 1$ mag for all galaxies, which is likely an incorrect assumption given what we find in \autoref{fig:sfrHa_AHa}.
Therefore we conclude there are more H$\alpha$ luminous galaxies than previously reported.
The impact of this finding on the SFRD is not important -- our SFRD is consistent with other works because SFR density contribution from the most H$\alpha$-luminous galaxies is small.
Larger surveys and spectroscopic follow-ups will allow us to further investigate this rare H$\alpha$-bright population.

The SFMS from our sample agrees well with previous results above the stellar mass completeness limit, $\log{(M/M_\odot)} > 9$.
Interestingly, we find there are many H$\alpha$ emitters with large sSFR below this mass limit. We show this population cannot be caused by scatter due to photometric errors.
We have measured the H$\alpha$/UV luminosity ratio of this population, which is a good indicator of starburst timescales. We find that the low-mass galaxies with large sSFR tend to have large ratios, suggesting that they experienced a shorter-timescale starburst recently.
In contrast, we find evidence that higher mass starburst galaxies have lower ratios and hence likely experienced a starbust mode for much longer timescales.

In this work, we have verified a method to derive H$\alpha$ fluxes from excesses in broad-band photometry, which enables us to efficiently construct large less-biased SFG sample.
This method also can be applied to other emission lines and/or other redshifts.
For example, the $J$ and $H$ medium-band filters of ZFOURGE can detect [OII] and H$\rm \beta$+[OIII] of galaxies in the same redshift range as H$\rm \alpha$ emitters in this work, respectively.

\section*{Acknowledgements}
We thank the anonymous reviewer for helpful comments that improved the paper. We thank Ivo Labb{\'e} for discussions related to this work. This work was supported by JSPS Overseas Challenge Program for Young Researchers and MEXT/JSPS KAKENHI Grant Number 15H02062, 18H03717, and 20H00171. LS acknowledges support from an Australian Research Council Discovery Project grant (DP190102448).